\documentclass{JHEP3}

\usepackage{graphicx,bbm}
\usepackage{amsmath,amsfonts,amssymb}
\usepackage{fleqn} 
\usepackage{mathrsfs}\usepackage{subfigure}
\usepackage{epsfig}\parskip 5pt plus 1pt
\usepackage[footnotesize]{caption}

{
}
\def\Tr{\mbox{Tr}\,}

\def\hbar{\hspace{0pt}\raisebox{1pt}{$-$} \hspace{-7pt} h}

\newcommand{\be}{\begin{equation}}
\newcommand{\ee}{\end{equation}}
\newcommand{\bd}{\begin{displaymath}}
\newcommand{\ed}{\end{displaymath}}
\newcommand{\bea}{\begin{eqnarray}}
\newcommand{\eea}{\end{eqnarray}}
\newcommand{\nn}{\nonumber}

\def\so10{$SO(10)$}

\title{Third Family Corrections to Quark and Lepton Mixing in SUSY
  Models with non-Abelian Family Symmetry}
\author{Stefan Antusch\\ 
Max-Planck-Institut f\"ur Physik (Werner-Heisenberg-Institut)\\
F\"ohringer Ring 6, D-80805 M\"unchen, Germany
\\ E-mail: \email{antusch@mppmu.mpg.de}} 
\author{Stephen
F. King\\ School of Physics and Astronomy, University of Southampton,
SO16 1BJ Southampton, United Kingdom\\ E-mail:
\email{sfk@hep.phys.soton.ac.uk}} \author{Michal Malinsk\'{y}\\ School
of Physics and Astronomy, University of Southampton, SO16 1BJ
Southampton, United Kingdom\\ E-mail:
\email{malinsky@phys.soton.ac.uk}}
\abstract{ We re-analyse the effect of corrections from canonical 
normalisation of kinetic terms on the quark and lepton mixing angles. This 
type of corrections emerges, for example, from effective higher-dimensional
K\"ahler potential operators in the context of locally supersymmetric
models of flavour. In contrast to previous studies we find that the
necessary procedure of redefining the fields in order to restore
canonically normalised kinetic terms, i.e.\ canonical normalisation,
can lead to significant corrections to the fermion mixing angles (as
determined from the superpotential). Such potentially large effects
are characteristic of flavour models based on non-Abelian family
symmetries, where some of the possible K\"ahler potential (and
superpotential) operators, in particular those associated with the
third family, are only mildly suppressed. We investigate
under which conditions the messenger sector of such flavour models
generates such K\"ahler potential operators for which the canonical
normalisation effects are sizeable, and under which conditions these
operators may be absent and canonical normalisation effects are
small. As explicit examples for potentially relevant CN effects, we
will discuss the corrections to the CKM matrix element $|V_{cb}|$ as
well as corrections to tri-bimaximal neutrino mixing.  }

\date{\today}

\begin{document}
%===========BODY - START===============================================
%%%%%%%%%%%%%%%%%%%%%%%%%%%%%%%%%%%%%%%%%%%%%%%%%%
%%%%%%%%%%%%%%%%%%%%%%%%%%%%%%%%%%%%%%%%%%%%%%%%%%
%%%%%%%%%%%%%%%%%%%%%%%%%%%%%%%%%%%%%%%%%%%%%%%%%%
%%%%%%%%%%%%%%%%%%%%%%%%%%%%%%%%%%%%%%%%%%%%%%%%%%
\newpage
\section{Introduction}
The flavour problem of the Standard Model (SM), i.e.\ the question of the origin of the observed fermion masses and mixings, is one of the deepest mysteries in particle physics. 
Since the discovery of the small neutrino masses and large lepton mixings has added new aspects to this problem, it has received much attention. 
In addition to adding to the flavour problem, the discoveries in the neutrino sector have also inspired new approaches towards its solution. As the precision of the neutrino data has improved, it has
become apparent that lepton mixing is consistent with the so called Tri-bimaximal (TB) mixing pattern \cite{HPS}, and many models attempt to reproduce this as a theoretical prediction
\cite{sumrule,Frampton:2004ud,Altarelli:2006kg,Ma:2007wu,deMedeirosVarzielas:2005ax,King:2006np,Harrison:2003aw,Chan:2007ng,Antusch:2007jd}. 
The essential starting point of many of such models
is to invoke a non-Abelian family symmetry which spans all
three families (like e.g.\ gauged $SO(3)$ or $SU(3)$,
or their discrete subgroups such as $A_4$ or $\Delta_{27}$ 
\cite{Altarelli:2006kg,Ma:2007wu,deMedeirosVarzielas:2005ax}) 
and which is subsequently spontaneously broken by extra Higgs scalars. 
In addition to explaining the large observed lepton mixing from an underlying tri-bimaximal pattern in the neutrino sector, models of this type can also accommodate the experimental data on quark masses and mixings.

If supersymmetry (SUSY) is discovered at the LHC, the presence of a spectrum of superparticle masses and their mixings and CP phases would add further aspects to the flavour problem. Models of flavour which are capable of addressing also these issues are typically formulated in the supergravity (SUGRA) framework.     
In this context, another intriguing aspect of some classes of the non-Abelian family symmetry models initially build to explain the (approximate) tri-bimaximal neutrino mixing, is that they can also provide a solution to this SUSY flavour and CP problem \cite{Antusch:2007re}. 
In these classes of flavour models, in the exact flavour symmetry limit the Yukawa couplings vanish and the matter sector K\"ahler metric becomes proportional to the unit matrix. Consequently, the soft SUSY breaking sfermion mass matrices are universal at leading order at high energies. Only after spontaneous flavour symmetry breaking by the vacuum expectation values (VEVs) $\langle\phi \rangle$ of the flavons, the Yukawa couplings are emerge from higher-dimensional operators involving flavon fields (and suppressed by powers of a messenger scale $M$). Their sizes can be expressed in terms of (powers of) expansion parameters $\varepsilon = \langle\phi \rangle/M$. In addition, after spontaneous flavour symmetry breaking effective higher-dimensional operators in the K\"ahler potential induce corrections to the universal sfermion mass matrices as well as  corrections to the kinetic terms (which have standard canonical form in leading order).

Before any flavour theory of this type can be reliably interpreted, field transformations must be performed in order to return the kinetic terms back to canonical form. These field transformations, however, in general lead to modifications of the Yukawa couplings and thus to the fermion masses and mixings (compared to their values extracted from the initial superpotential). This rather technical but necessary procedure, to which we will refer to as canonical normalisation (CN) in the following, has been discussed in \cite{Leurer:1993gy} and more recently in \cite{Antusch:2007re,King:2003xq,King:2004tx}. 
Although these possible effects have been to some extent addressed in
the previous works \cite{King:2003xq} and \cite{King:2004tx}, the conclusion of these studies has been that the effects on the mixing angles are too small to be relevant. 

The motivation for re-visiting the effects of CN in this study is the observation that in classes of theories that predict TB mixing, especially those based on non-Abelian family symmetries spanning all three families of SM matter, certain K\"ahler potential operators can occur which are only very mildly suppressed. These operators lead to non-universal entries in the CN transformation matrices of order $\varepsilon_3^2$, where $\varepsilon_3^2 =|\langle\phi_3 \rangle|/M_3$, with typically $\varepsilon_3 \approx 0.5$ \cite{Antusch:2007re}. The reason for the appearance of this rather large ``expansion parameter'' is that the large third generation Yukawa couplings (in particular $y_{t}$) must originate from a effective vertex containing an insertion of (at least) one flavon field (here called $\phi_{3}$). To accommodate for example a large $y_t$, typically a rather large parameter $\varepsilon_3 \approx 0.5$ is introduced. 
For such large non-universality in the CN transformations, one can anticipate that their effects on fermion masses and mixings cannot be neglected anymore.
In a recent study \cite{Antusch:2007ib} focusing on the corrections to lepton sector mixing, it has been highlighted that such large third family wave-function corrections have to be included when comparing the model prediction of TB mixing in the neutrino sector with precision data of future neutrino oscillation facilities \cite{Group:2007kx}.  

The main purpose of this paper is therefore to analyse in detail the possible impact of such potentially large CN effects on the quark and lepton mixing angles, encoded in the Cabibbo-Kobayashi-Maskawa (CKM) and Pontecorvo-Maki-Nakagawa-Sakata (PMNS) mixing matrices respectively. Another main question we will investigate in detail in this study is under which conditions the messenger sector of such flavour models generates these types of mildly suppressed K\"ahler potential operators and under which conditions these operators may be absent such that the corrections from canonical normalisation may be small. As explicit examples for potentially relevant CN effects, we will discuss the corrections to the CKM matrix element $|V_{cb}|$ as well as corrections to tri-bimaximal neutrino mixing. Regarding the tri-bimaximal mixing example, we go beyond the analysis of \cite{Antusch:2007ib} by considering CN corrections in a realistic class of SU(3) flavour symmetry models and by providing additional details regarding the derivation of the CN results and of the procedure of combining CN corrections with corrections from renormalisation group (RG) running and Cabibbo-like charged lepton mixing contributions (i.e.\ regarding the there proposed stable mixing sum rule).

The paper is organised as follows: In the subsequent
section we shall comment on some  generalities of the canonical
normalisation procedure focusing on ambiguities in the
definition of the canonical normalisation transformation. We develop a
perturbative technique to deal with effects of the canonical
redefinition of fields in the Yukawa sector focusing in particular on
its impact on the CKM and PMNS mixing parameters. Sections 3 and 4 are
then devoted to a set of examples of the CN effects in the quark and
lepton sectors respectively. 
In section 5 we present a more in-depth discussion of
the expected magnitude of the CN corrections in a particular class of realistic 
SUSY flavour models based on non-Abelian family symmetry.
In section 6 we compare the CN corrections to other typically relevant 
corrections. Section 7 concludes the paper. 
Some technical aspects of the discussion in the main body of the paper and a 
specification of the used conventions can be found in the Appendices.

%%%%%%%%%%%%%%%%%%%%%%%%%%%%%%%%%%%%%%%%%%%%%%%%%%
%====================================================================
\section{The K\"ahler potential and effects of canonical normalisation}
%====================================================================
Whenever the K\"ahler potential of a given SUSY model is nontrivial there are extra effects coming from the canonical normalisation procedure bringing the generic kinetic terms 
 \begin{eqnarray}
{\cal L}^{\tilde{f}}_\mathrm{kin} &=& 
\partial_\mu \tilde{Q}_{i\alpha}^{*} (K_{Q})_{ij} \partial^\mu\tilde{Q}_{\alpha j}+
\partial_\mu \tilde{u}^{c*}_{i} (K_{u})_{ij} \partial^\mu\tilde{u}^{c}_{j}+
\partial_\mu \tilde{d}^{c*}_{i} (K_{d})_{ij} \partial^\mu\tilde{d}^{c}_{j}+\ldots,
\nonumber \\
%& & \nonumber \\
{\cal L}^{f}_\mathrm{kin} &=& 
\overline{Q}_{i\alpha} (K_{Q})_{ij}i\gamma^{\mu} \partial_\mu {Q}_{\alpha j}+
\overline{u^{c}}_{i} (K_{u})_{ij}i\gamma^{\mu} \partial_\mu {u^{c}}_{j}+
\overline{d^{c}}_{i} (K_{d})_{ij}i\gamma^{\mu} \partial_\mu {d^{c}}_{j}+\ldots,
\end{eqnarray}
(where $(K_{f})_{ij}$ denotes the K\"ahler metric for the given scalar $\tilde{f}$, $\tilde{f}^{c}$ and fermionic ${f}$, ${f}^{c}$ degrees of freedom) into the canonical form ${K}_{f} =\delta_{ij}(\tilde f_{\mathrm{can}}^{*})_{i}(\tilde f_{\mathrm{can}})_{j}$ and  $\delta_{ij}(\overline f_{\mathrm{can}})_{i}(f_{\mathrm{can}})_{j}$ respectively. 

In a wide class of non-Abelian flavour models, the dominant contributions to $K_{f,f^{c}}$'s come from insertions of the flavon field associated to the third family Yukawas{\footnote{In the flavour models based on $SU(3)$ family symmetry, the third family Yukawa couplings are governed by operators of the type $\tfrac{1}{M^{2}}(f.\phi_{3})(f^{c}.\phi_{3})H$ exploiting the triplet nature of matter of both chiralities $f$, $f^{c}$, while in the $SO(3)$ theories the structure of the leading order operators typically looks like $\tfrac{1}{M}(f.\phi_{3}) f^{c}_{3}H$ due to the singlet nature of the right-handed spinors. The dot product corresponds to the simplest bilinear invariants of the flavour symmetry under consideration.}} (usually denoted by $\phi_{3}$) yielding
\be\label{simpleKahler}
K_{f,f^{c}}\approx \overline{f}_{i}\left[k_{0}^{f}\delta_{ij}+k_{3}^{f}\frac{1}{M_{\psi}^{2}}\langle\phi_{3}^{\dagger}\rangle_{i}\langle\phi_{3}\rangle_{j}\right]f_{j}+\overline{f^{c}}_{i}\left[k_{0}^{f^{c}}\delta_{ij}+k_{3}^{f^{c}}\frac{1}{M_{\chi}^{2}}\langle\phi_{3}^{\dagger}\rangle_{i}\langle\phi_{3}\rangle_{j}\right]f_{j}^{c}\;,
\ee 
where $k_{0,3}^{f,f^{c}}$ are real constants and $M_{\psi}$ and $M_{\chi}$ denote masses of the messenger fields relevant for the left- and right-chirality matter sectors respectively. If the flavon VEV $\langle\phi_{3}\rangle$ is comparable to either $M_{\psi}$ and/or $M_{\chi}$, one can expect a potentially large deviations from the leading order universality (governed by the $\delta_{ij}$ factors above) in the relevant part of the K\"ahler metric. 

It is important that one can hardly trim all these contributions to zero simultaneously by fiddling around with the messenger masses because there is no symmetry that could prevent every Yukawa sector relevant messenger from entering either $K_{f}$ or $K_{f^{c}}$. This\footnote{As we shall see, it is namely the left-handed sector (i.e.\ non-universalities in $K_{f}$) that could have large effects on quark and lepton mixing in the charged currents.}, however, depends strongly on the character of the Yukawa operators, and can, in turn, single out a class of particular ``K\"ahler-corrections-safe'' flavour models; for more detailed discussion see section \ref{models}.
 
%---------------------------------------------------------
\subsection{Definitions and ambiguities\label{sectionDefinitionsAmiguities}}
%---------------------------------------------------------
Canonical normalisation consists in redefining the defining basis fields ${f}$ and ${f}^{c}$ so that the original (for instance scalar sector) kinetic terms  ${\cal L}_\mathrm{kin}=\partial{\hat f}^{\dagger}K_{f}\partial\hat{f}+\partial\hat{f}^{c\dagger}K_{f^{c}}\partial\hat{f}^{c}$ receive the canonical form ${\cal L}^{\mathrm{can}}_\mathrm{kin}= \partial   f^{\dagger}\partial  f+\partial  f^{c\dagger}\partial  f^{c}$. This is achieved\footnote{In what follows, we shall focus on the left-chiral matter sector, i.e.\ we shall often give the results only for $K_{f}$; the relevant formulae for $K_{f^{c}}$ can be obtained  upon replacing all the super-(sub-)scripts $f\to f^{c}$.} by transforming the defining superfields by $\hat{f}\to P_{f}^{-1}\hat{f}\equiv f_{\mathrm{can}}$ where $P_{f}$ is a matrix bringing the relevant K\"ahler metric $K_{f}$ into the diagonal form %(unless necessary from now on we shall suppress all the flavour indices)
: 
\be\label{Pmatrices}
P_{f}^{\dagger}K_{f} P_{f}=\mathbbm{1},\quad \mathrm{i.e.}\quad K_{f}=P_{f}^{\dagger-1}P_{f}^{-1}.
\ee
This can be easily done in two steps:
\begin{itemize}
\item{First, one can always diagonalise the Hermitean K\"ahler metric by means of a unitary transformation $U_{f}K_{f} U_{f}^{\dagger}=K_{f}^{D}$ where $K^{D}_{f}$ is a real diagonal matrix.
}
\item{Second, a diagonal rescaling by $\sqrt{K^{D}_{f}}^{-1}$ from both sides of $U_{f}K_{f} U_{f}^{\dagger}=K_{f}^{D}$ drives its RHS to unity:  
$\left(\sqrt{K^{D}_{f}}\right)^{-1}U_{f}K_{f} U_{f}^{\dagger}\left(\sqrt{K^{D}_{f}}\right)^{-1}=\mathbbm{1}$. Moreover, one can multiply this formula by any unitary matrix from the left (and its inverse from the right) with no effect on this unity matrix.}
\end{itemize}
Thus, the most generic form of $P_{f}$ reads 
\be\label{genericP}
P_{f}=U_{f}^{\dagger}\left({\sqrt{K^{D}_{f}}}\right)^{-1}\tilde U_{f},
\ee  
with the freedom to choose the unitary $\tilde{U}_{f}$ matrix arbitrarily. Thus, one can for instance have $P_{f}$ Hermitean by choosing $\tilde{U}_{f}=U_{f}$ or exploit this freedom to bring the $P_{f}$ into a triangular form as e.g.\ in \cite{King:2004tx}.

\paragraph {Note on notation:}\mbox{}\\
In what follows, whenever appropriate we use hats to denote quantities in the defining basis (i.e.\ before canonical normalisation) while the unhatted symbols correspond to their physical counterparts, i.e.\ to quantities after the CN effects were already taken into account. 
%---------------------------------------------------------
\subsection{Effect of canonical normalisation on the Yukawa couplings}
%---------------------------------------------------------
\paragraph {Lepton sector:}\mbox{}\\
Suppose the original charged lepton and the light Majorana neutrino mass matrices are diagonalised\footnote{The situation in the quark and lepton sectors is different: since the quark sector diagonalisation transformation is {\it bi-unitary}, one can always absorb the would-be phases of the diagonal entries of $M_{f}^{D}$  by a suitable redefinition of $V_{L}^{f}$ and $V_{R}^{f}$ and get rid of all but one CP phase in the CKM matrix. This is not possible for Majorana neutrinos as there is only one unitary matrix in the relevant formula. This, in turn, gives rise to extra phase factors associated to PMNS mixing - the Majorana phases.} by means of the biunitary transformations
$\hat V_{L}^{l}\hat M_{l}\hat V_{R}^{l\dagger}=\hat M_{l}^{D}$ and 
$
\hat V_{L}^{\nu}\hat m_{\nu}\hat V_{L}^{\nu T}=\hat m_{\nu}^{D}
$
so that the lepton mixing matrix (before canonical normalisation) obeys $\hat U_{PMNS}=\hat V_{L}^{l}\hat V_{L}^{\nu\dagger}$. The effect of canonical normalisation on $\hat M_{l}$ and\footnote{Note that the $P_{\nu^{c}}$ actually does not enter the effective light neutrino matrix because it cancels among the right-handed components of the neutrino Yukawas and the inverse of the Majorana mass matrix in the seesaw formula $\hat m_{\nu}= M_{\nu}^{D}M_{M}^{-1}( M_{\nu}^{D})^{T}$ \cite{seesaw}.}  $\hat M_{\nu}$:
\be\label{cannormmatrices}
\hat M_{l}\to P_{L}^{T}\hat M_{l}P_{e^{c}}\equiv M_{l} 
\qquad \mathrm{and}\qquad
\hat m_{\nu}\to P_{L}^{T}\hat m_{\nu}P_{L}\equiv m_{\nu} 
\ee
induces a relevant change on $\hat V_{L,R}^{l}\to V_{L,R}^{l}$ and $\hat V_{L}^{\nu}\to V_{L}^{\nu}$ so that 
$
V_{L}^{l}M_{l}V_{R}^{l\dagger}=M_{l}^{D}$
and  
$
V_{L}^{\nu}m_{\nu}V_{L}^{\nu T}=m_{\nu}^{D}
$ is fulfilled, i.e.:
\be\label{basicrelations}
V_{L}^{l}P_{L}^{T}\hat M_{l}P_{e^{c}}V_{R}^{l\dagger}=M_{l}^{D}
\qquad \mathrm{and}\qquad  
V_{L}^{\nu}P_{L}^{T}\hat m_{\nu}P_{L}V_{L}^{\nu T}=m_{\nu}^{D}
\ee
should be satisfied. Then the {\it physical} lepton mixing matrix obeys (up to the rephasing bringing it into the standard PDG form \cite{Yao:2006px})
$
U_{PMNS}=V_{L}^{l}V_{L}^{\nu\dagger}
$.

\paragraph {Quark sector:}\mbox{}\\
The reasoning for the quark sector goes along the same lines as above - the original basis up and down-type Yukawa matrices $\hat M_{u,d}$ diagonalisable by biunitary transformations
$\hat V_{L}^{u}\hat M_{u}\hat V_{R}^{u\dagger}=\hat M_{u}^{D}$
and  
$\hat V_{L}^{d}\hat M_{d}\hat V_{L}^{d\dagger}=\hat M_{d}^{D}$ (leading to $\hat V_{CKM}=\hat V_{L}^{u}\hat V_{L}^{d\dagger}$ before canonical normalisation)
change upon canonical normalisation into
\be\label{cannormmatricesq}
\hat M_{u}\to P_{Q}^{T}\hat M_{u}P_{u^{c}}\equiv M_{u} 
\qquad \mathrm{and}\qquad
\hat M_{d}\to P_{Q}^{T}\hat M_{d}P_{d^{c}}\equiv M_{d} 
\ee
inducing a change on $\hat V_{L,R}^{u}\to V_{L,R}^{u}$ and $\hat V_{L,R}^{d}\to V_{L,R}^{d}$ so that 
\be\label{basicrelationsq}
V_{L}^{u}P_{Q}^{T}\hat M_{u}P_{u^{c}}V_{R}^{u\dagger}=M_{u}^{D}
\qquad \mathrm{and}\qquad  
V_{L}^{d}P_{Q}^{T}\hat M_{d}P_{d^{c}}V_{R}^{d\dagger}=M_{d}^{D}.
\ee
The physical CKM matrix then obeys 
$
V_{CKM}=V_{L}^{u}V_{L}^{d\dagger}$.

%---------------------------------------------------------
\subsection*{Irrelevance of $\tilde{U}_{f}$ matrices}
%---------------------------------------------------------
It is easy to see that the arbitrary $\tilde U_{f}$ matrices  in the definition of $P_{f}$ do not play any role in either the mixing matrices $U_{PMNS}$, $V_{CKM}$ or the physical spectra. Indeed, under any unitary change $\tilde{U}'_{f}$ of the relevant $\tilde{U}_{f}$ matrices in the definition (\ref{genericP}), i.e.\ $P_{f}\to P_{f}\tilde{U}'_{f}$, the effects in (\ref{basicrelations}) can be absorbed into redefinitions $V_{L}^{l}\to V_{L}^{l}\tilde{U}'{}^{\dagger}_{L}$ and $V_{L}^{\nu}\to V_{L}^{\nu}\tilde{U}'{}^{\dagger}_{L} $ so that (\ref{basicrelations}) remains unaffected. However, $\tilde{U}'_{L}$ cancels in $U_{PMNS}$ and the physical spectra remain intact, because the would-be effects of $\tilde{U}_{L}$ and  $\tilde{U}_{e^{c}}$ matrices in (\ref{basicrelations}) can be absorbed into the biunitary transformation revealing the spectrum of the charged lepton Yukawa matrix. Similarly, one can justify the irrelevance of the particular choice of $P_{Q}$ and $P_{u^{c},d^{c}}$ for $V_{CKM}$ and the quark sector spectra.

%---------------------------------------------------------
\subsection*{Exploiting the freedom in definition of $P_{f,f^{c}}$}
%---------------------------------------------------------
Thus, one can exploit the freedom in choosing $\tilde U_{f,f^{c}}$ matrices in the definition of $P_{f,f^{c}}$ to simplify the structure of (\ref{genericP}) so that the $P_{f,f^{c}}$-factors are particularly easy to handle. The convenient choice is indicated by the fact that even if the original K\"ahler metric is just a slight perturbation of the unity matrix $\mathbbm{1}$ (up to an irrelevant overall normalisation $k_{0}^{f}$), the diagonalisation matrix $U_{f}$ in $P_{f}=U_{f}^{\dagger} ({\sqrt{K^{D}_{f}}})^{-1}\tilde U_{f}$ could still be large. The intention to write  $({\sqrt{K^{D}_{f}}})^{-1}$ as $(\mathbbm{1}+\Delta K^{D}_{f})/{\sqrt{k_{0}^{f}}}$, exploiting the limited departure of the K\"ahler metric spectrum from unity, then gives $P_{f}=(U_{f}^{\dagger} \tilde{U}_{f}+U_{f}^{\dagger}\Delta K^{D}_{f}\tilde{U}_{f})/{\sqrt{k_{0}^{f}}}$ that could be brought to a particularly convenient form for $\tilde U_{f}=U_{f}$, and we can benefit from $P_{f}=(\mathbbm{1}+U_{f}^{\dagger}\Delta K^{D}_{f}U_{f})/{\sqrt{k_{0}^{f}}}$, i.e.\ hermiticity of $\Delta P_{f}\equiv U_{f}^{\dagger}\Delta K^{D}_{f}U_{f}$ and simplicity of $P_{f}=(\mathbbm{1}+\Delta P_{f})/{\sqrt{k_{0}^{f}}}$.
%---------------------------------------------------------
\subsection{Perturbative prescription for the physical rotation matrices\label{sectionperturbative}}
%---------------------------------------------------------
For Hermitean $P_{L, e^{c}}$ and $P_{Q,u^{c},d^{c}}$  and $P_{f, f^{c}}=(\mathbbm{1}+\Delta P_{f, f^{c}})/\sqrt{k_{0}^{f,f^{c}}}$ (assuming a small departure of $K_{f}$ and $K_{f^{c}}$ from unity), one obtains from (\ref{basicrelations}):
\bea\label{basicrelations2}
V_{L}^{l}(\mathbbm{1}+\Delta P_{L}^{T})\hat M_{l}(\mathbbm{1}+\Delta P_{e^{c}})V_{R}^{l\dagger} & = & {\sqrt{k_{0}^{L}k_{0}^{e^{c}}}} M_{l}^{D}\nn \;,  \\
V_{L}^{\nu}(\mathbbm{1}+\Delta P_{L}^{T})\hat m_{\nu}(\mathbbm{1}+\Delta P_{L})V_{L}^{\nu T}&=&  k_{0}^{L} m_{\nu}^{D}
\eea
for the lepton sector and from (\ref{basicrelationsq}):
\bea\label{basicrelations2q}
V_{L}^{u}(\mathbbm{1}+\Delta P_{Q}^{T})\hat M_{u}(\mathbbm{1}+\Delta P_{u^{c}})V_{R}^{u\dagger} & = & \sqrt{k_{0}^{Q}k_{0}^{u^{c}}}M_{u}^{D}\nn  \;,  \\
V_{L}^{d}(\mathbbm{1}+\Delta P_{Q}^{T})\hat M_{d}(\mathbbm{1}+\Delta P_{d^{c}})V_{R}^{d\dagger} & = & \sqrt{k_{0}^{Q}k_{0}^{d^{c}}}M_{d}^{D}
\eea
for the quarks.
If all the (high-scale) physical spectra are sufficiently hierarchical\footnote{This is certainly true for all charged matter fermion spectra; for neutrinos we shall stick to the hierarchical spectrum case from now on.}, the smallness of $\Delta P_{f,f^{c}}$ factors ensures only small differences between the hatted and un-hatted diagonalisation matrices, i.e.
\be\label{mixingchange}
V_{L,R}^{f}=W_{L,R}^{f} \hat V_{L,R}^{f}\;,
\ee
where $W_{L,R}^{f}$ are small unitary rotations in the unity neighbourhood (up to a phase ambiguity to be discussed later): 
\be\label{Wmatrices}
W_{L,R}^{f}=\mathbbm{1} + i \Delta W_{L,R}^{f}\;,
\ee
with $\Delta W_{L,R}^{f}$ denoting their Hermitean generators. One can disentangle the left-handed and right-handed rotations in formulae (\ref{basicrelations2}) and (\ref{basicrelations2q}) by  considering $M_{f}M_{f}^{\dagger}$:
\bea
W_{L}^{l}\hat V_{L}^{l}(\mathbbm{1}+\Delta P_{L}^{T})\hat M_{l}(\mathbbm{1}+2\Delta P_{e^{c}})\hat M_{l}^{\dagger}(\mathbbm{1}+\Delta P_{L}^{*}) \hat V_{L}^{l\dagger}W_{L}^{l\dagger}& = & k_{0}^{L}k_{0}^{e^{c}}M_{l}^{D}M_{l}^{D
\dagger} \nn  \;,  \\
W_{L}^{u}\hat V_{L}^{u}(\mathbbm{1}+\Delta P_{Q}^{T})\hat M_{u}(\mathbbm{1}+2\Delta P_{u^{c}})\hat M_{u}^{\dagger}(\mathbbm{1}+\Delta P_{Q}^{*}) \hat V_{L}^{u\dagger}W_{L}^{u\dagger}& = & {k_{0}^{Q}k_{0}^{u^{c}}} M_{u}^{D}M_{u}^{D
\dagger}\label{basicrelations3} \;,  \\
W_{L}^{d}\hat V_{L}^{d}(\mathbbm{1}+\Delta P_{Q}^{T})\hat M_{d}(\mathbbm{1}+2\Delta P_{d^{c}})\hat M_{d}^{\dagger}(\mathbbm{1}+\Delta P_{Q}^{*}) \hat V_{L}^{d\dagger}W_{L}^{d\dagger}& = & {k_{0}^{Q}k_{0}^{d^{c}}} M_{d}^{D}M_{d}^{D
\dagger} \;,  \nn
\eea
which yields (from the three complex off-diagonal zero conditions) at the leading order: 
\bea
\!\!\!\!\!\!\!\!\!\!\!(\Delta W_L^{l})_{ij,i\neq j}&=&\frac{i}{\hat m_{j}^{l2}-\hat m_{i}^{l2}}
\left[
(\hat m_{i}^{l2}+\hat m_{j}^{l2})\left(\hat V_{L}^{l}\Delta P_{L}^{T}\hat V_{L}^{l\dagger}\right)_{ij}+
2\hat m_{i}^{l}\hat m_{j}^{l}\left(\hat V_{R}^{l}\Delta P_{e^{c}}\hat V_{R}^{l\dagger}\right)_{ij}
\right]\nn  \label{DeltaWl}  \;,  \\
\!\!\!\!\!\!\!\!\!\!\!(\Delta W_L^{u})_{ij,i\neq j}&=&\frac{i}{\hat m_{j}^{u2}-\hat m_{i}^{u2}}
\left[
(\hat m_{i}^{u2}+\hat m_{j}^{u2})\left(\hat V_{L}^{u}\Delta P_{Q}^{T}\hat V_{L}^{u\dagger}\right)_{ij}+
2\hat m_{i}^{u}\hat m_{j}^{u}\left(\hat V_{R}^{u}\Delta P_{u^{c}}\hat V_{R}^{u\dagger}\right)_{ij}
\right]\nn \label{DeltaWu} \;,  \\
\!\!\!\!\!\!\!\!\!\!\!(\Delta W_L^{d})_{ij,i\neq j}&=&\frac{i}{\hat m_{j}^{d2}-\hat m_{i}^{d2}}
\left[
(\hat m_{i}^{d2}+\hat m_{j}^{d2})\left(\hat V_{L}^{d}\Delta P_{Q}^{T}\hat V_{L}^{d\dagger}\right)_{ij}+
2\hat m_{i}^{d}\hat m_{j}^{d}\left(\hat V_{R}^{d}\Delta P_{d^{c}}\hat V_{R}^{d\dagger}\right)_{ij}
\right]\nn \label{DeltaWd} \;,  \\
\eea
where the eigenvalues $\hat m_{i}^{f2}$ of the original $\hat M_{f}$ matrices can be at the leading order identified with the physical charged fermion masses and the overall normalisation factors $k_{0}^{f,f^{c}}$ drop.
Similarly, the neutrino sector corrections obey (replacing $\hat M_{l}\to \hat m_{\nu}$, $V_{L}^{l}\to V_{L}^{\nu}$, $V_{R}^{l}\to V_{L}^{\nu*}$ and $\Delta P_{e^{c}}\to \Delta P_{L}$ in the first formula above)
\bea
\!\!\!\!\!\!\!\!\!\!\!(\Delta W_L^{\nu})_{ij,i\neq j}=\frac{i}{\hat m_{j}^{\nu2}-\hat m_{i}^{\nu2}}
\left[
(\hat m_{i}^{\nu2}+\hat m_{j}^{\nu2})\left(\hat V_{L}^{\nu}\Delta P_{L}^{T}\hat V_{L}^{\nu\dagger}\right)_{ij}+
2\hat m_{i}^{\nu}\hat m_{j}^{\nu}\left(\hat V_{L}^{\nu}\Delta P_{L}^{T}\hat V_{L}^{\nu\dagger}\right)_{ji}
\right]. \nn \\ \; \label{DeltaWnu}
\eea
Due to the assumed hierarchy in the physical spectra, the first terms tend to dominate over the second (thus screening the ambiguity in the unknown structure of the right-handed rotations in the charged sector) and we shall often neglect the latter.

Notice that formulae (\ref{DeltaWd}), (\ref{DeltaWnu}) provide only the off-diagonal entries of $\Delta W^{f}_{L}$'s. However, this reflects the three phase ambiguity in defining the diagonalisation matrices $W$ by means of relations like (\ref{basicrelations3}). Thus, it is not surprising that three parameters in $\Delta W^{f}_{L}$'s remain unconstrained and can be in principle chosen arbitrarily with the only constraint coming from the required perturbativity of the $W^{f}_{L}$ matrices (\ref{Wmatrices}). For simplicity, we shall put the diagonal entries of all $W^{f}_{L}$'s to zero keeping in mind the possible need for ``standard'' rephasing of the physical lepton mixing matrix. Another reason is that in the real case $W$ become orthogonal and thus generated by antisymmetric purely imaginary $\Delta W^{f}_{L}$'s. Thus, the $\Delta W^{f}_{L}$ matrices can be without loss of generality chosen in the form:
\be\label{corrections}
\Delta W^{f}_{L}=\begin{pmatrix}
0 & \Delta W^{f}_{L12} & \Delta W^{f}_{L13} \\
\Delta W_{L12}^{f*} & 0 & \Delta W^{f}_{L23} \\
 \Delta W_{L13}^{f*} &  \Delta W_{L23}^{f*} & 0
\end{pmatrix},
\ee
with the off-diagonal entries given by formulae (\ref{DeltaWl}) and (\ref{DeltaWnu}). 
With this at hand one can write the physical\,\footnote{From now on we shall always choose the free phases in $\hat{V}_{L,R}^{f}$ (i.e.\ work in a particular basis) so that the ${\hat V}_{CKM}$ and $\hat {U}_{PMNS}$ matrices are in their 'standard' form \cite{Yao:2006px}. This, however, need not be the case after the CN corrections are taken into account and we shall comment on the phases later.} quark and lepton mixing matrices $V_{CKM}$ and $U_{PMNS}$ in term of the original ones $\hat V_{CKM}$  and $\hat U_{PMNS}$ as:
\bea
U_{PMNS}&=&(\mathbbm{1}+i\Delta W_{L}^{l})\hat U_{PMNS}(\mathbbm{1}-i\Delta W_{L}^{\nu\dagger})=\hat U_{PMNS}+\Delta U_{PMNS} \nn \; ,\\
V_{CKM}&=&(\mathbbm{1}+i\Delta W_{L}^{u})\hat V_{CKM}(\mathbbm{1}-i\Delta W_{L}^{d\dagger})=\hat V_{CKM}+\Delta V_{CKM}\;,\nn
\eea
with\footnote{Although $\Delta W_{L,R}^{f}$ are by definition Hermitean, we shall often keep the dagger in formulae like (\ref{MNSchange}), (\ref{CKMcorrs}) to help reader's orientation in the text.} 
\bea\label{MNSchange}
\Delta U_{PMNS}&=&i\left(\Delta W_{L}^{l}\hat U_{PMNS}-\hat U_{PMNS}\Delta W_{L}^{\nu\dagger}\right)+\ldots \; ,\\
\Delta V_{CKM}&=&i\left(\Delta W_{L}^{u}\hat V_{CKM}- \hat V_{CKM}\Delta W_{L}^{d\dagger}\right)+\ldots\label{CKMcorrs}\;.
\eea
Recall that in a particular model, all the ingredients are actually at hand - one can easily diagonalise the Hermitean K\"ahler metric to get the (conventionally) Hermitean $P^{-1}_{f}$ factors (and from there $\Delta P_{f}$'s) and the various $\hat V_{L,R}^{f}$ matrices in (\ref{DeltaWl}), (\ref{DeltaWnu}) can be inferred in the same manner from the underlying model Yukawa couplings.
%===========================================================
\section{Canonical normalisation corrections to quark sector mixing}
%===========================================================
To illustrate the importance of CN corrections for the quark sector mixing we discuss as an example the dominant K\"ahler corrections to the $V_{cb}$ CKM entry in the class of potentially realistic $SU(3)$ setting with a large third family expansion parameter. Such large third family expansion parameter appears, e.g., in the models discussed in \cite{Antusch:2007re,deMedeirosVarzielas:2006fc,Malinsky:2007pf}.

%-----------------------------------------------------------------------
\subsection{Corrections to $V_{cb}$ in classes of $SU(3)$ flavour models}
%-----------------------------------------------------------------------
For simplicity reasons, we shall focus on a {\it real} 2$\times$2 case for the two heavy states only for a quasi-diagonal LH quark sector  K\"ahler metric  along the lines
of  \cite{deMedeirosVarzielas:2006fc} discussed in great detail in e.g.\ \cite{Antusch:2007re,Malinsky:2007pf}. Assuming that the expansion parameters in the K\"ahler sector coincide with those relevant for the superpotential (c.f. section \ref{models} for a detailed discussion of this point), the relevant piece of the matter sector K\"ahler metric can be written as\footnote{For sake of simplicity, we have chosen a particular shape of $K_{Q}$ so that the numerical factors are simple.} 
\be
K_{Q}=k_{0}^{Q}\begin{pmatrix}
1& \varepsilon^{2}\\
\varepsilon^{2} & 1+\varepsilon_{3}^{2}
\end{pmatrix},
\ee
which is diagonalised by means of 
$
U_{Q}K_{Q}U_{Q}^{\dagger}= K_{Q}^{D}$,
where:
\be
U_{Q}\approx \begin{pmatrix}
1& -\tfrac{\varepsilon^{2}}{\varepsilon_{3}^{2}}\\
\tfrac{\varepsilon^{2}}{\varepsilon_{3}^{2}} & 1
\end{pmatrix}\quad \mathrm{and}\quad K_{Q}^{D}\approx k_{0}^{Q}\mathrm{diag}(1,1+\varepsilon_{3}^{2}) \;.
\ee
Adopting the Hermitean convention $\tilde U_{Q}=U_{Q}$, i.e.\ $P_{Q}=U_{Q}^{\dagger}(K_{Q}^{D})^{-\tfrac{1}{2}}U_{Q}/{\sqrt{k_{0}^{Q}}}$, one obtains\footnote{Note that for $\tilde{U}_{Q}=\mathbbm{1}$ one receives $P_{Q}\approx
\frac{1}{\sqrt{k_{0}^{Q}}}\begin{pmatrix}
1& \tfrac{\varepsilon^{2}}{\varepsilon_{3}^{2}}\\
-\tfrac{\varepsilon^{2}}{\varepsilon_{3}^{2}} &1-\tfrac{\varepsilon_{3}^{2}}{2}
\end{pmatrix}$ instead with enhanced off-diagonal terms with respect to (\ref{PQ}).
As it was pointed out in \cite{King:2003xq}, such a $P_{Q}$ matrix can induce a potentially large deviation of the physical Yukawa matrices from their defining basis structure. However, as far as physical observables such as the CKM mixings are concerned, the individual relatively large 23 rotations arising in such case in both up and down sectors act against each other and leave only a subleading effect, which becomes almost trivial to infer upon adopting $\tilde{U}_{Q}=U_{Q}$. 

In short, the ``Hermitean'' convention for $P_{Q}$'s adopted here does not induce large fake corrections to the off-diagonal Yukawa couplings and the corresponding $V_{L}^{u,d}$ matrices.
}
\be\label{PQ}
P_{Q}
\approx
\frac{1}{\sqrt{k_{0}^{Q}}}\begin{pmatrix}
1& -\tfrac{\varepsilon^{2}}{2}\\
-\tfrac{\varepsilon^{2}}{2} &1-\tfrac{\varepsilon_{3}^{2}}{2}
\end{pmatrix}\;,
\ee 
and the physical Yukawas obey (at the leading order)
\bea
\hat Y_{u}&\approx &\begin{pmatrix}
\varepsilon^{2}& \varepsilon^{2}\\
\varepsilon^{2} & \varepsilon_{3}^{2}
\end{pmatrix}
\to Y_{u}=(P_{1Q})^{T}\hat Y_{u}P_{u^{c}}\approx
\frac{1}{\sqrt{k_{0}^{Q}}}\begin{pmatrix}
\varepsilon^{2}&\varepsilon^{2}-\tfrac{1}{2}\varepsilon^{2}_{3}\varepsilon^{2}\\
\varepsilon^{2} & \varepsilon_{3}^{2}
\end{pmatrix}P_{u^{c}}\nn \;,\\
\hat Y_{d}&\approx &\begin{pmatrix}
\overline\varepsilon^{2}& \overline\varepsilon^{2}\\
\overline\varepsilon^{2} & \varepsilon_{3}^{2}
\end{pmatrix}
\to Y_{d}=(P_{1Q})^{T}\hat Y_{d}P_{d^{c}}\approx
\frac{1}{\sqrt{k_{0}^{Q}}}\begin{pmatrix}
\overline\varepsilon^{2}& \overline\varepsilon^{2}-\tfrac{1}{2}\varepsilon^{2}_{3}\varepsilon^{2}\\
\overline\varepsilon^{2} & \varepsilon_{3}^{2}
\end{pmatrix}P_{d^{c}}\;,
\eea
indicating non-negligible {\it additive} leading order corrections to 23 rotations in $V_{L}^{u,d}$, that, however, cancel at the leading order in the CKM mixing matrix. 

The net effect eventually emerges from the next to leading order ratio of the 23 and 33 entries and can be readily obtained from the perturbative prescription (\ref{CKMcorrs}) together with (\ref{DeltaWd}) provided:
\be
\Delta P_{Q}=-\frac{1}{2}\begin{pmatrix}
0&\varepsilon^{2}\\
\varepsilon^{2}& \varepsilon_{3}^{2}
\end{pmatrix}, \quad 
\hat V_{L}^{u}\approx 
\begin{pmatrix}
1&-\tfrac{\varepsilon^{2}}{\varepsilon_{3}^{2}}\\
\tfrac{\varepsilon^{2}}{\varepsilon_{3}^{2}}& 1
\end{pmatrix}, \quad 
\hat V_{L}^{d}\approx 
\begin{pmatrix}
1&-\tfrac{\overline\varepsilon^{2}}{\varepsilon_{3}^{2}}\\
\tfrac{\overline\varepsilon^{2}}{\varepsilon_{3}^{2}}& 1
\end{pmatrix},
\ee
(giving
$\hat V_{CKM}\approx
\begin{pmatrix}
1& \tfrac{\overline\varepsilon^{2}-\varepsilon^{2}}{\varepsilon_{3}^{2}}\\
-\tfrac{\overline\varepsilon^{2}-\varepsilon^{2}}{\varepsilon_{3}^{2}} & 1
\end{pmatrix}
$
and thus $\hat V_{cb}\approx \tfrac{\overline\varepsilon^{2}-\varepsilon^{2}}{\varepsilon_{3}^{2}}$ before canonical normalisation). This  
yields at leading order $\Delta W_{L}^{u}\approx 0$ and from (\ref{DeltaWd})
$
\Delta W_{L}^{d}\approx i\begin{pmatrix}
0& \tfrac{\overline\varepsilon^{2}-\varepsilon^{2}}{2}\\
-\tfrac{\overline\varepsilon^{2}-\varepsilon^{2}}{2} & 0
\end{pmatrix}
$. Therefore, (\ref{CKMcorrs}) leads to:
\bea
\Delta V_{CKM}\approx -i \hat V_{CKM} \Delta W_{L}^{d\dagger}\approx
\begin{pmatrix}
0& \tfrac{\overline\varepsilon^{2}-\varepsilon^{2}}{2}\\
\tfrac{\varepsilon^{2}-\overline\varepsilon^{2}}{2} & 0
\end{pmatrix},
\eea
so the CKM matrix changes after canonical normalisation into:
\bea
V_{CKM}\approx 
\begin{pmatrix}
1& \tfrac{\overline\varepsilon^{2}-\varepsilon^{2}}{\varepsilon_{3}^{2}}\left(1+\tfrac{1}{2}\varepsilon_{3}^{2}\right)\\
\tfrac{\varepsilon^{2}-\overline\varepsilon^{2}}{\varepsilon_{3}^{2}}\left(1+\tfrac{1}{2}\varepsilon_{3}^{2}\right) & 1
\end{pmatrix}.
\eea
The physical value of the 23 quark-sector mixing is then modified to: 
\be\label{Vtbcorrection}
V_{cb}= \hat V_{cb}\left(1+\frac{1}{2}\varepsilon_{3}^{2}\right)+\ldots.
\ee
As anticipated, there is a relatively large {\it multiplicative} correction due to the presence of the large expansion parameter associated to the third family canonical normalisation corrections, that was not appreciated in \cite{King:2004tx}.

%%%%%%%%%%%%%%%%%%%%%%%%%%%%%%%%%%%%%%%%%
\section{Canonical normalisation corrections to lepton sector mixing\label{Kahlercorrections}}
%%%%%%%%%%%%%%%%%%%%%%%%%%%%%%%%%%%%%%%%%
In order to study the effects of K\"ahler corrections to a generic bi-large lepton sector mixing, one can not avoid the first generation anymore. Thus, in what follows, we consider the full $3\times3$ structure of the relevant mixing matrices as well as the matter sector K\"ahler metric. 
%-----------------------------------------------------------------------
\subsection{Corrections due to third family canonical rescaling}
%-----------------------------------------------------------------------

Though the generic shape of the relevant piece of the K\"ahler metric (i.e.\ namely $K_{L}$ as far the lepton sector is concerned\footnote{Recall that $\Delta P_{e^{c}}$ is screened in  (\ref{DeltaWd}) at the leading order and corrections due to $P_{\nu^{c}}$ entirely cancel in the seesaw formula, c.f. (\ref{cannormmatrices}).}) is rather complicated, in realistic cases one can expect the dominant effects coming from the leading non-universal contribution  (\ref{simpleKahler}) governed by $\langle\phi_{3}\rangle$. 
Thus, we shall first focus on the simplified setting where only the entries due to (\ref{simpleKahler}) are taken into account. Later on (in section \ref{realistic}), we shall compare the results obtained here with the full-fledged potentially realistic $SU(3)$ model analysis to reveal that this is indeed a very accurate approximation. 

In the present case, the lepton sector K\"ahler metric is given at leading order by:
\be\label{defeta0}
K_{L}=k_{0}^{L}\left(\mathbbm{1}+\frac{k_{3}^{L}}{k_{0}^{L}}\frac{\langle\phi_{3}^{\dagger}\rangle\langle\phi_{3}\rangle}{M_{K}^{2}}\right)+\ldots\;,
\ee
with $k_i^{L}$ denoting the relevant ${\cal O}(1)$ Wilson coefficients in (\ref{simpleKahler}), while $M_{K}$ stands for a generic K\"ahler sector messenger mass. In models where the 33 Yukawa entries are (at least partly) generated by means of $SU(2)_{L}$-doublet messengers  (that in turn enter also the K\"ahler potential) $M_{K}$ is around the scale of the relevant Yukawa-sector-active messengers (denoted by $\chi_{i}$ in section \ref{models}) and $\frac{\langle\phi_{3}\rangle}{M_{K}}$ is of the order of the Yukawa sector parameter $\varepsilon_{3}$. 

At the leading order, the lepton sector K\"ahler metric can be written in a matrix form:
\be\label{Kahler33}
K_{L}\approx{k_{0}^{L}}\left[\mathbbm{1}+
\begin{pmatrix}
0 & 0 & 0 \\
0 & 0 & 0 \\
0 & 0 & \eta^\mathrm{K}
\end{pmatrix}\right]\; \quad \mathrm{where}\quad \eta^\mathrm{K}\equiv \frac{k_{3}^{L}}{k_{0}^{L}}\frac{|\langle\phi_{3}\rangle|^{2}}{M_{K}^{2}}.
\ee   
Therefore, the $P_{L}$ matrix is just:  
\be\label{DeltaP}
P_{L} =\frac{1}{\sqrt{k_{0}^{L}}}\left[\mathbbm{1}-
\frac{1}{2}\begin{pmatrix}
0 & 0 & 0 \\
0 & 0 & 0 \\
0 & 0 & \eta^\mathrm{K}
\end{pmatrix}\right]\;,\qquad \mathrm{and\;\;thus}\qquad
\Delta P_{L} =-
\frac{1}{2}\begin{pmatrix}
0 & 0 & 0 \\
0 & 0 & 0 \\
0 & 0 & \eta^\mathrm{K}
\end{pmatrix}.
\ee   
In the next sub-section, we shall consider the canonical normalisation corrections models of (nearly) tri-bimaximal mixing in the lepton sector. 

%=======================================================
\subsection{Canonical normalisation corrections to tri-bimaximal neutrino mixing\label{simplestsetup}}
%=======================================================
As an example for the impact of the potentially large third family CN corrections on lepton mixing, let us consider their effects on the pattern of exact tri-bimaximal lepton mixing. In many classes of flavour models this pattern of tri-bimaximal mixing emerges as a prediction of the neutrino sector \cite{King:2006hn}. These models are inspired by the proximity of the present neutrino oscillation data on PMNS matrix to the  tri-bimaximal mixing matrix \`a la Harisson-Perkins-Scott \cite{Harrison:2003aw}, which has the form
\be
U_{TB}=
\begin{pmatrix}
\sqrt{\tfrac{2}{3}} & \tfrac{1}{\sqrt{3}} & 0 \\
- \tfrac{1}{\sqrt{6}}  & \tfrac{1}{\sqrt{3}} &  \tfrac{1}{\sqrt{2}} \\
 \tfrac{1}{\sqrt{6}} & - \tfrac{1}{\sqrt{3}} &  \tfrac{1}{\sqrt{2}}
\end{pmatrix}.\,P_{M}\quad 
\mathrm{with}\quad
P_{M}=
\begin{pmatrix}
e^{i\tfrac{\alpha_{1}}{2}} & 0& 0\\
0& e^{i\tfrac{\alpha_{2}}{2}}  & 0\\
0& 0& 1
\end{pmatrix},
\ee
where $P_{M}$ is (so far) experimentally undetermined diagonal matrix encoding the two observable Majorana phase differences.

We shall first focus on the simplest setting and assume that the lepton mixing generated by the underlying family symmetry happens to be exactly tri-bimaximal (in the defining basis) , i.e.\ $\hat U_{PMNS}=\hat V_{L}^{l}\hat V_{L}^{\nu\dagger}= U_{TB}$ and comes entirely from the neutrino sector \cite{King:2006hn}, i.e.\ $\hat V_{L,R}^{l}\approx \mathbbm{1}$ while $\hat V_{L}^{\nu\dagger}\approx U_{TB}$.
In the canonical basis, $\hat U_{PMNS}=U_{TB}$ changes along (\ref{MNSchange}) yielding: 
\be
U_{PMNS}=U_{TB}+i\left(\Delta W_{L}^{l}U_{TB}- U_{TB}\Delta W_{L}^{\nu\dagger}\right)+\ldots\;,
\ee
and the correction matrices are given by (\ref{corrections}) provided (\ref{DeltaWl}), (\ref{DeltaWnu}). 
Taking into account the screening of the second terms $\propto 2\hat m_{i}\hat m_{j}/(\hat m_{i}^{2}-\hat m_{j}^{2})$ in formulae (\ref{DeltaWl}) and (\ref{DeltaWnu}) in case of the hierarchical neutrino spectrum, one obtains\footnote{From now on, we shall always assume that the defining basis masses $\hat m_{i}^{f}$ coincide at the leading order with the corresponding physical quantities.}: 
\bea\label{Wsimplest}
(\Delta W_L^{l})_{ij,i\neq j}& \approx & \frac{i(\hat m_{i}^{l2}+\hat m_{j}^{l2})}{\hat m_{j}^{l2}-\hat m_{i}^{l2}}
\left(\Delta P_{L}^{T}\right)_{ij}=0 \;,\nn
\\
(\Delta W_L^{\nu})_{ij,i<j}& \approx &\frac{i(\hat m_{i}^{\nu2}+\hat m_{j}^{\nu2})}{\hat m_{j}^{\nu2}-\hat m_{i}^{\nu2}}
\left(U_{TB}^{\dagger}\Delta P_{L}^{T}U_{TB}\right)_{ij}\approx i \left(U_{TB}^{\dagger}\Delta P_{L}^{T}U_{TB}\right)_{ij}+\ldots \;.\label{firstterm}
\eea
that yields at the leading order
\be\label{universalproportionality}
\Delta (U_{TB})_{ij,i<j} \approx - (U_{TB})_{i\underline{j}}\left(U_{TB}^{\dagger}\Delta P_{L}^{T}U_{TB}\right)_{\underline{j}\underline{j}}\quad \mathrm{(no\;summation\;over\;} j \mathrm{)}\;.
\ee
Remarkably enough, the corrections to the three matrix elements under consideration are (at the leading order) proportional to their values, c.f. formula (\ref{universalproportionality}) and thus, in particular, the canonical normalisation corrections to the reactor angle are cancelled by the 13 zero of $U_{TB}$. Second, the Majorana phases are irrelevant for the second bracket on the LHS of formula (\ref{universalproportionality}) and enter only through the first term. Thus, the phase structure of the correction is identical to the phase structure of the original matrix element and there is no need for an additional rephasing.

\FIGURE{
\centering
$\ensuremath{\vcenter{\hbox{\includegraphics[width=10cm]{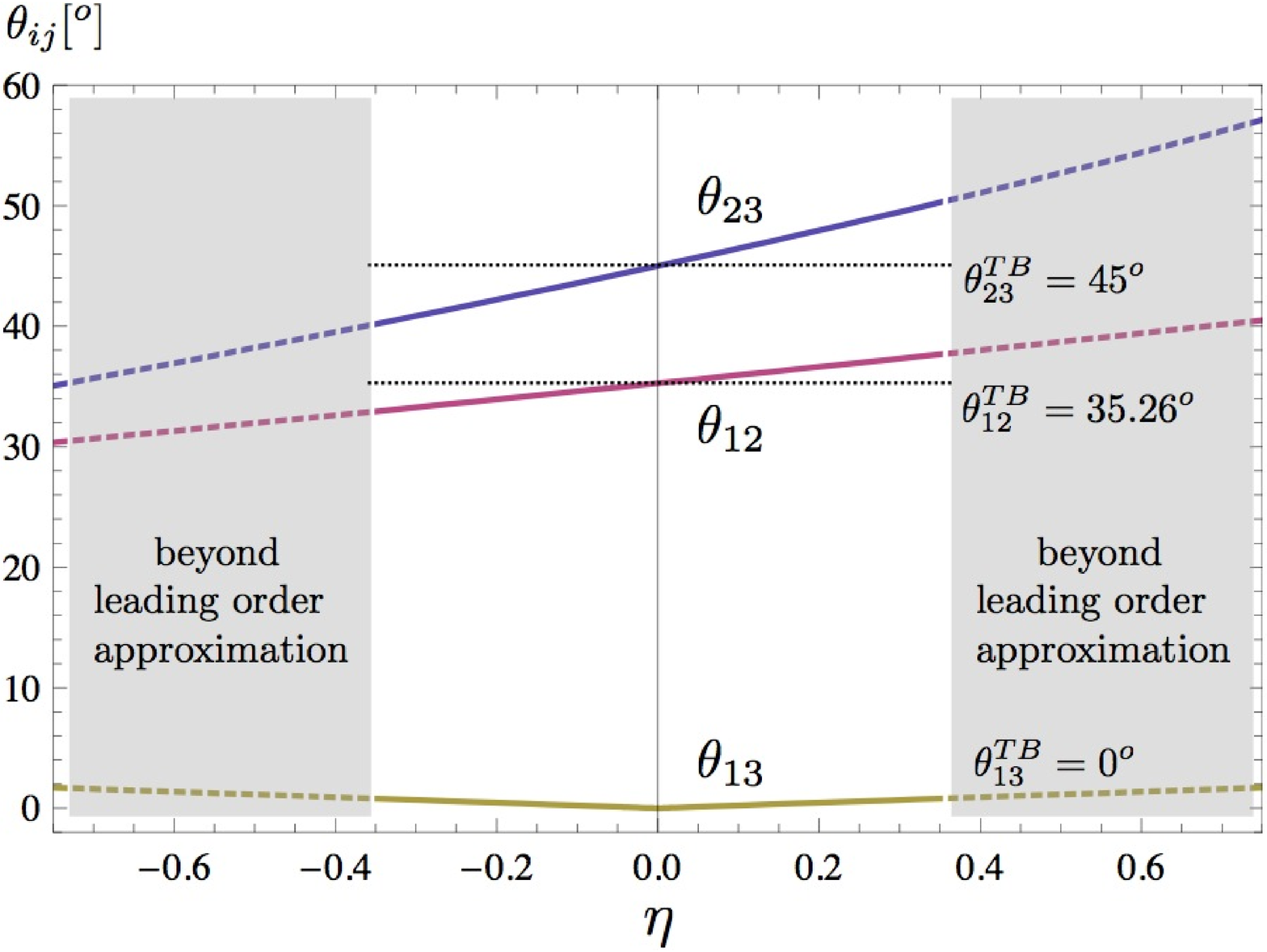}}}}$ 
 \caption{\label{fig:corrections-theory} 
Corrections to the tri-bimaximal neutrino mixing due to canonical normalisation (taken from \cite{Antusch:2007ib}). The $\eta$-parameter stands for $\eta=\eta^{\mathrm{K}}$ from (\ref{Kahler33}) for the canonical normalisation effects, $\eta=\eta^{\mathrm{RG}}$ from (\ref{Eq:etaRG}) for the leading order running effects, or $\eta=\eta^{\mathrm{K}}+\eta^{\mathrm{RG}}$ if both CN and RG corrections are taken into account, c.f. (\ref{etatotal}). The shaded regions correspond to  values of $|\eta|$ for which one can expect deviations from the leading-order perturbative results. The reactor mixing in this setup is seen to be rather stable with respect to the considered third family corrections.}
}

Numerically, this leads for example to $(\Delta U_{TB})_{12}\approx \eta^\mathrm{K}\frac{1}{6\sqrt{3}}$, $(\Delta U_{TB})_{23}\approx \eta^\mathrm{K}\frac{1}{4\sqrt{2}}$ and $(\Delta U_{TB})_{13}\approx 0$ (up to irrelevant phase factors).
The zero in the 13 correction, however, emerges only from the first term in the approximation (\ref{firstterm}) and gets lifted at the next-to-leading level. Indeed, employing the full-featured formula (\ref{DeltaWnu}) one recovers (for hierarchical case):
\bea
\Delta (U_{PMNS})_{13}&=&2\left[\frac{\hat m^{\nu}_{2}}{\hat m^{\nu}_{3}}(U_{TB})_{12}(U_{TB}^{\dagger})_{23}+\frac{\hat m^{\nu}_{1}}{\hat m^{\nu}_{3}}(U_{TB})_{11}(U_{TB}^{\dagger})_{13}\right]\!\!(\Delta P_{L})_{33}(U_{TB})_{33}\nn\\
&=& -\frac{1}{3\sqrt{2}}\left(\frac{\hat m^{\nu}_{2}}{\hat m^{\nu}_{3}}e^{i\tfrac{\hat \alpha_{2}}{2}}-\frac{\hat m^{\nu}_{1}}{\hat m^{\nu}_{3}}e^{i\tfrac{\hat \alpha_{1}}{2}}\right)\eta^\mathrm{K} \;,\label{subleading}
\eea
where the two phase-factors reflect the Majorana nature of the light neutrino masses. 
The last formula finally yields (assuming the first term in the bracket dominates):
\be
\theta_{13}\approx  |\eta^\mathrm{K}|\frac{1}{3\sqrt{2}}\sqrt{\frac{\Delta m^{2}_{\odot}}{\Delta m^{2}_{A}}}\approx 4\times 10^{-2}|\eta^\mathrm{K}|\;.
\ee
All together, this gives at the leading order:
\be
U_{PMNS}\approx
\begin{pmatrix}
\sqrt{\tfrac{2}{3}}-\eta^\mathrm{K}\tfrac{1}{6\sqrt{6}} & \tfrac{1}{\sqrt{3}}+\eta^\mathrm{K}\tfrac{1}{6\sqrt{3}} & 4\times 10^{-2}|\eta^\mathrm{K}|e^{-i\delta} \\
- \tfrac{1}{\sqrt{6}}+\eta^\mathrm{K}\tfrac{1}{12\sqrt{6}}  & \tfrac{1}{\sqrt{3}}-\eta^\mathrm{K}\tfrac{2}{6\sqrt{3}} &  \tfrac{1}{\sqrt{2}}+\eta^\mathrm{K}\tfrac{1}{4\sqrt{2}} \\
 \tfrac{1}{\sqrt{6}}+\eta^\mathrm{K}\tfrac{5}{12\sqrt{6}} & - \tfrac{1}{\sqrt{3}}-\eta^\mathrm{K}\tfrac{1}{6\sqrt{3}} &  \tfrac{1}{\sqrt{2}}-\eta^\mathrm{K}\tfrac{1}{4\sqrt{2}}
\end{pmatrix}. \, P_{M}\;.\label{correctionTBcase}
\ee

\FIGURE{
\centering
$\ensuremath{\vcenter{\hbox{\includegraphics[width=10cm]{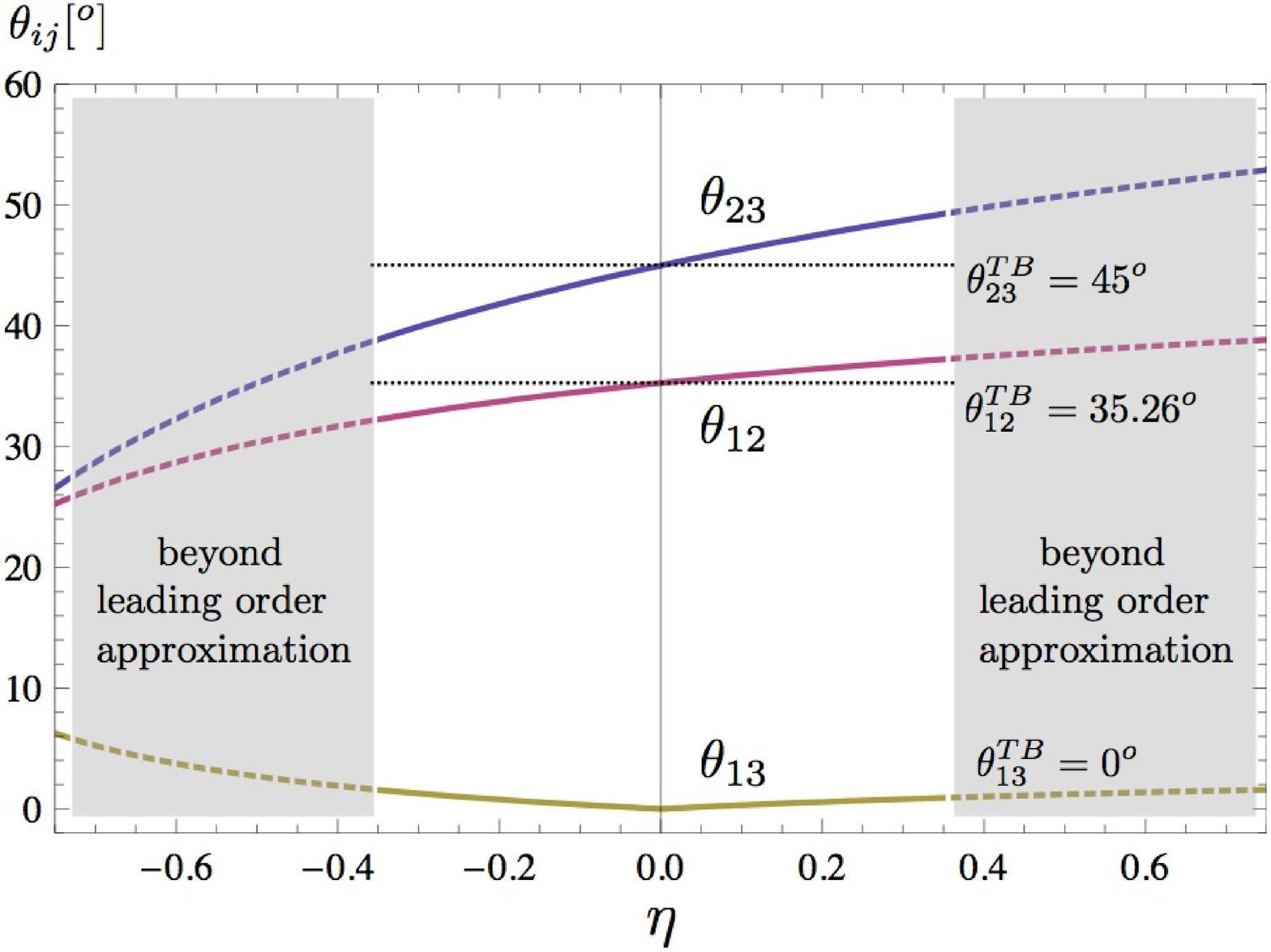}}}}$ 
 \caption{\label{fig:corrections} 
Corrections to the tri-bimaximal neutrino mixing (faint dashed lines) from canonical 
normalisation of the kinetic terms as a function of 
$\eta =\eta^{\mathrm{K}}\equiv\varepsilon_3^2\frac{k^L_3}{k^L_0}$ in a potentially realistic  $SU(3)_{f}$ flavour model \cite{Antusch:2007re} discussed in detail in section \ref{realistic}. The displayed curves correspond to the leading order approximate results given by formulae (\ref{SU3results}). Remarkably enough, these results coincide for small $\eta$ with those obtained by perturbative methods in the simplified setup discussed is section \ref{simplestsetup}, thus demonstrating the crucial role played by the dominant 33-sector K\"ahler correction.}
}

It can be easily checked that $U_{PMNS}$ is unitary up to ${\cal O}(\eta^{K2})$ terms. 
Exploiting the parametrisation of \cite{King:2007pr} one gets: 
\be
s_{12}=\frac{1}{\sqrt{3}}(1+s),\quad 
s_{23}=\frac{1}{\sqrt{2}}(1+a),\quad 
s_{13}=\frac{1}{\sqrt{2}}r\;,
\ee
and (comparing to (\ref{correctionTBcase})) the `TB-deviation' parameters\footnote{Recall that the 13 mixing can be always without loss of generality made positive by a suitable redefinition of the lepton sector Dirac CP phase.}  read:
\be\label{rsa}
r\approx 6\times 10^{-2} |\eta^\mathrm{K}|,\quad s=\frac{\eta^\mathrm{K}}{6}\quad \mathrm{and}\quad a=\frac{\eta^\mathrm{K}}{4}\;.
\ee 
We see, in particular, that $\theta_{13}$ is rather stable and that the atmospheric mixing is changing faster than the solar ($a=\eta^\mathrm{K}/4$ while $s=\eta^\mathrm{K}/6$), c.f. the shape of 
curves depicted in Fig.\ref{fig:corrections-theory}.

%%%%%%%%%%%%%%%%%%%%%%%%%%%%%%%%%%%%%%%%%%%%%%
\section{Canonical normalisation corrections in potentially realistic models}
%%%%%%%%%%%%%%%%%%%%%%%%%%%%%%%%%%%%%%%%%%%%%%
As discussed in the Introduction, 
it has been pointed out that the observed close-to
tri-bimaximal lepton mixing, along with the main features of the quark
and charged lepton sector observables, can be understood in frameworks with non-Abelian family symmetry $({\cal F})$, that is spontaneously broken by the VEVs of  
three flavons $\phi_3 , \phi_{23} , \phi_{123}$ (transforming as triplets under ${\cal F}$) pointing in particular directions in the family space. These flavon fields give rise to the Yukawa  
operators of the shape (in the case of $SU(3)$ family symmetry, dropping superfield hats) 
\cite{deMedeirosVarzielas:2005ax}:
\be
\label{TBYukawa} 
\!\!\!\!\!\!\!W_{Y}\approx
f^{i}f^{cj}
\frac{H}{M_{f}^{2}}\left[y_1^f(\phi_{123})_{i}(\phi_{23})_{j}\!+\!y_2^f(\phi_{23})_{i}(\phi_{123})_{j}\!+\!y_3^f(\phi_{3})_{i}(\phi_{3})_{j}\!+\!{y_{4}^f}(\phi_{23})_{i}(\phi_{23})_{j}\right]\!.
\ee
This approach requires the vacuum alignment in the $SU(3)$ space of the form: 
$\phi_3  \sim (0,0,1)$, $\phi_{23}  \sim (0,1,1)$,
$\phi_{123}  \sim (1,1,1)$, up to phases. 

Note also that there is in principle at least two distinct types of
messengers entering the formula (\ref{TBYukawa}), in particular
those transmitting the $SU(2)_{L}$ doublet nature of $f=Q, L$ to the
Higgs VEV insertion point (for definiteness let's call them
$\chi_{Q,L}$), and the $SU(2)_{L}$-singlets propagating further the
remaining $SU(3)_{c}\otimes U(1)_{Y}$ quantum numbers to
$f^{c}=u^{c}$, $d^{c}$, $e^{c}$ and $\nu^{c}$ (to be called
$\chi_{u,d,e,\nu}$), c.f. Figure \ref{threestructures}. However, for sake of simplicity, we shall use a
generic symbol $M_{f}$ for both these classes and come back to this
distinction only upon getting to physical implications.
Later in this section we shall address the question of topology of
the underlying messenger sector Feynman graphs giving rise to the
operators under consideration. We shall also discuss the
relationship between these messengers and those which appear
in the K\"ahler potential.
%=======================================================
\subsection{Corrections to tri-bimaximal mixing in a class of $SU(3)_{f}$ flavour models\label{realistic}}
%=======================================================

As an example, let us focus on the canonical normalisation corrections to the tri-bimaximal neutrino mixing in the classes of $SU(3)_{f}$ 
flavour models considered in \cite{Antusch:2007re}.
Taking into account the irrelevance of the canonical transformation $P_{\nu^{c}}$ in the see-saw formula for the light neutrino masses, the quantity of our main interest is the leading order K\"ahler metric for the 
lepton doublets $K_{L}$ obeying:
\begin{eqnarray}\label{SU3KL}
K_{L} \approx k_0^{L} \mathbbm{1}
+
\left(\begin{array}{lll}
\varepsilon_K^4 k^L_{1}&  \varepsilon_K^4 k^L_{1}  e^{i \phi_1}& \varepsilon_K^4 k^L_{1}  e^{i \phi_2}\\
.&  \varepsilon_K^2 k^L_{2} & \varepsilon_K^2 k^L_{2}e^{i \phi_3}\\
.& .&   \varepsilon_{K3}^2k^L_3
\end{array}\right)+\ldots \;,
%, \\
\end{eqnarray}
where the subscript $K$ in the expansion parameters $\varepsilon_{K}$ and $\varepsilon_{K3}$ indicates that the K\"ahler metric messenger masses $M_K$ (entering through e.g.\ $\varepsilon_{K3} = |\langle \phi_3\rangle|/M_K$) may differ from those relevant for the Yukawa sector (\ref{TBYukawa}) and the dotted terms in (\ref{SU3KL}) can be reconstructed from hermiticity.
The $P_{L}^{-1}$ matrix is obtained\footnote{Recall that relation (\ref{Pmatrices}) fixes the $P$-matrices 
only up to a global unitary transformation; as before we adopt 
the convention $\tilde U_{L}=U_{L}$so that $P$'s are Hermitean, c.f. section \ref{sectionDefinitionsAmiguities}.} to leading order in $\varepsilon_{K}, \varepsilon_{K3}$ as:
\be
P_{L}^{-1} = \begin{pmatrix}\label{Pf}
\sqrt{k_{0}^L}  + \frac{\varepsilon_K^4 k^L_{1}}{2\sqrt{k_0^L}}&  \frac{\varepsilon_K^4 k^L_{1}  e^{i \phi_1}}{2\sqrt{k_0^L}}& \frac{\varepsilon_K^4 k^L_{1}  e^{i \phi_2}}{\sqrt{k_0^L} + \sqrt{k_0^L + k^L_3 \varepsilon_{K3}^2}}\\
\frac{\varepsilon_K^4 k^L_{1}  e^{-i \phi_1}}{2\sqrt{k_0^L}}& \sqrt{k_0^L} + \frac{\varepsilon_K^2 k^L_{2}}{2\sqrt{k_0^L}} & \frac{\varepsilon_K^2 k^L_{2} e^{i \phi_3}}{\sqrt{k_0^L} + \sqrt{k_0^L + k^L_3 \varepsilon_{K3}^2}}\\
\frac{\varepsilon_K^4 k^L_{1}  e^{-i \phi_2}}{\sqrt{k_0^L} + \sqrt{k_0^L + k^L_3 \varepsilon_{K3}^2}}& \frac{\varepsilon_K^2 k^L_{2}e^{-i \phi_3}}{\sqrt{k_0^L} + \sqrt{k_0^L + k^L_3 \varepsilon_{K3}^2}}& \sqrt{k_0^L + k^L_3 \varepsilon_{K3}^2} 
\end{pmatrix}+\ldots \;.
\ee
Notice that due to the relatively large $\varepsilon_{3K}\sim 0.5$, the naive factorisation 
$P= \frac{1}{\sqrt{k_{0}}}(\mathbbm{1}+\Delta P)$ (with $|\Delta P|=-\tfrac{1}{2}\Delta K_{L}\ll \mathbbm{1}$ for $K_{L}\equiv k_{0}^{L}(\mathbbm{1}+\Delta K_{L})$) is 
violated in the third family due to higher power $\varepsilon_{K3}$-effects.

\paragraph{Charged lepton sector:}\mbox{}\\
Inspecting the charged lepton Yukawa matrices in this class of models (c.f.\ \cite{Antusch:2007re}) before and after canonical normalisation, it can be seen that the charged lepton mixing angles themselves as well as 
the CN corrections are small. 
Therefore we can still treat the charged lepton mixing angles as only (CKM-like) small corrections to the 
neutrino sector dominated $U_{PMNS}$ and we shall (first) focus on the neutrino sector.

\paragraph{Neutrino sector:}\mbox{}\\
In the class of models under consideration, the Majorana mass matrix $\hat M_{M}$ originates from 
operators which involve factors like  
$f^{ci}f^{cj}(\phi_{23})_{i}(\phi_{23})_{j}$ and 
$f^{ci}f^{cj}(\phi_{123})_{i}(\phi_{123})_{j}$.
The relevant matrix structures read \cite{deMedeirosVarzielas:2005ax}:
\begin{equation}
\label{Eq:Matrices}
\!\!\!\!\!\!\!\! \hat Y^{\nu}=
\left( \begin{array}{ccc}
0 & B & C_1\\
A & Be^{i \phi_1}+Ae^{i \phi_1} & C_2\\
Ae^{i \phi_3} & Be^{i \phi_2}+Ae^{i (\phi_1 + \phi_3 )} & C_3
\end{array}
\right)
,\;
\hat M_{M}=
\left( \begin{array}{ccc}
M_A & M_A e^{i \phi_1} & 0    \\
M_A e^{i \phi_1} & M_A e^{2i \phi_1}+M_B & 0    \\
0 & 0 & M_C
\end{array}
\right)\!,
\end{equation}
where the real positive entries in $\hat M_{M}$ satisfy $M_A<M_B<M_C$. 
In terms of the expansion parameters \cite{Antusch:2007re}, the neutrino Yukawa matrix is given by
\be
\hat Y^\nu = \begin{pmatrix}
0 & \varepsilon^3 y_1& \varepsilon^3 y_1 e^{i \phi_3}\\
\varepsilon^3 y_2& \varepsilon^3 (y_1 e^{i \phi_1} + y_2 e^{i \phi_1}) & \varepsilon^3 (y_1 e^{i (\phi_1 + \phi_3)} + y_2 e^{i \phi_2})\\
\varepsilon^3 y_2 e^{i \phi_3}& \varepsilon^3 (y_1 e^{i \phi_2} + y_2 e^{i (\phi_1+\phi_3)}) & y_3 \varepsilon_3^2
\end{pmatrix} +\ldots\;,
\ee
which matches Eq.~(\ref{Eq:Matrices}) with $A=y_2^{\nu}\varepsilon^3$ and $B=y_1^{\nu}\varepsilon^3$, $\varepsilon\sim 0.05$. 
The CN transformation (\ref{cannormmatrices}) then  yields (since $P_{\nu^c}$ drops off the seesaw formula for the light Majorana neutrinos, we are free to choose for simplicity $P_{\nu^{c}}=\mathbbm{1}$) $M_{M}=\hat M_{M}$ and:
\be
Y^\nu = \begin{pmatrix}
{\cal O}(\varepsilon^7) & \frac{\varepsilon^3 y_1}{\sqrt{k^L_0}}& \frac{\varepsilon^3 y_1 e^{i \phi_3}}{\sqrt{k^L_0}}\\
\frac{\varepsilon^3 y_2}{\sqrt{k^L_0}}& \frac{\varepsilon^3 (y_1 e^{i \phi_1} + y_2 e^{i
\phi_1})}{\sqrt{k^L_0}} & {\cal O}(\varepsilon^2)\\
\frac{\varepsilon^3 y_2 e^{i \phi_3}}{\sqrt{k^L_0}\sqrt{1+\varepsilon_{K3}^2\frac{k^L_3}{k^L_0}}}& \frac{\varepsilon^3 (y_1 e^{i \phi_2} + y_2 e^{i (\phi_1+\phi_3)})}{\sqrt{k^L_0}\sqrt{1+\varepsilon_{K3}^2\frac{k^L_3}{k^L_0}}} & \frac{y_3 \varepsilon_3^2}{\sqrt{k^L_0}\sqrt{1+\varepsilon_{K3}^2\frac{k^L_3}{k^L_0}}}
\end{pmatrix} +\ldots\;.
\ee
In order to extract the mixing angles analytically from these matrices, it is convenient to
transform $Y^{{\nu}}$ and $M_{M}$ by means of a suitable non-singular matrix $S$ \cite{King:2006hn,Antusch:2007re}:
\be
Y^{{\nu}}  \rightarrow  Y'{}^{{\nu}}=Y^{{\nu}}
\,S^{-1},\ \ 
M  \rightarrow M'=
{S^T}^{-1} \,M\,S^{-1}, \ \ 
M^{-1}  \rightarrow M'{}^{-1}=
{S}\,M^{-1}\,S^{T}\;,
\label{S}
\ee
(which again leaves the neutrino mass matrix invariant) to the case of a diagonal $M_{M} = \mbox{diag}(M_A,M_B,M_C)$ which corresponds to: 
\be
Y'{}^\nu = \begin{pmatrix}\label{Eq:Ynuprime}
{\cal O}(\varepsilon^7) & \frac{\varepsilon^3 y_1}{\sqrt{k^L_0}}& \frac{\varepsilon^3 y_1 e^{i \phi_3}}{\sqrt{k^L_0}}\\
\frac{\varepsilon^3 y_2}{\sqrt{k^L_0}}& \frac{\varepsilon^3 y_1 e^{i \phi_1}}{\sqrt{k^L_0}} & {\cal O}(\varepsilon^2)\\
\frac{\varepsilon^3 y_2 e^{i \phi_3}}{\sqrt{k^L_0}\sqrt{1+\varepsilon_{K3}^2\frac{k^L_3}{k^L_0}}}& \frac{\varepsilon^3 y_1 e^{i \phi_2}}{\sqrt{k^L_0}\sqrt{1+\varepsilon_{K3}^2\frac{k^L_3}{k^L_0}}} & \frac{y_3 \varepsilon_3^2}{\sqrt{k^L_0}\sqrt{1+\varepsilon_{K3}^2\frac{k^L_3}{k^L_0}}}
\end{pmatrix} +\ldots \;.
\ee

\paragraph{Formulae for the corrected neutrino mixing angles:}\mbox{}\\
Since the last transformation brought the neutrino Yukawa and Majorana matrices into a particular form along the lines of the Sequential Dominance setting \cite{King:1998jw}, from Eq.~(\ref{Eq:Ynuprime}) we can directly read off the mixing
angles (imposing $\phi_2 - \phi_1 = \phi_3 - \pi$) at leading order in $m_2/m_3$ (making use of the generic formulae given in \cite{King:1998jw}):
\be
\!\!\! \tan\theta_{23}^\nu \approx \sqrt{1+ \eta^\mathrm{K}} \; ,\;\;
\label{SU3results}
\tan\theta_{12}^\nu \approx \frac{1}{{c_{23}^\nu + \frac{s_{23}^\nu}{\sqrt{1+ \eta^\mathrm{K}}}}}\; ,\;\;
\theta_{13}^\nu \approx \frac{m_2}{m_3}  \frac{(s_{12}^\nu)^2|\eta^\mathrm{K}|}{\sqrt{(1+\eta^\mathrm{K})(2+\eta^\mathrm{K})}} \;.\nn
\ee
with $\eta^\mathrm{K} \equiv k_3^L \varepsilon_{K3}^2 / k_0^L$. The $\eta^\mathrm{K}$-behaviour of these relations is illustrated in Fig.~\ref{fig:corrections}. 
We see that this independent calculation confirms the findings of the previous sections for small $\eta^\mathrm{K}$.
To give a quantitative example we may take $\varepsilon_{K3} = 0.5$ and set the ${\cal O}(1)$ coefficients
$k^L_3$ and $k^L_0$ to $1$ yielding $\eta^\mathrm{K}=0.25$ and
$\theta^\nu_{23} = 50.8^\circ$, 
$\theta^\nu_{12} = 38.3^\circ$ and
$\theta^\nu_{13} = 1.1^\circ$, 
compared to the tri-bimaximal neutrino mixing predictions
$\hat\theta^\nu_{23} = 45^\circ,\hat\theta^\nu_{12} = 35.26^\circ$ and $\hat\theta^\nu_{13} = 0^\circ$ before
canonical normalisation.

%=========================================================================
\subsection{Heavy $SU(2)_{L}$ doublet messengers \& natural $\eta^\mathrm{K}$ suppression\label{models}}
%=========================================================================
Since the naive estimate of the CN effects  above leads to non-negligible deviations from the TB-mixing in the lepton sector (in particular for relatively large $|\eta^\mathrm{K}|$), let us sketch in brief the prospects of getting $k_{3}^{L}/k_{0}^{L}$ (and thus $\eta^\mathrm{K}$) naturally suppressed in the class of popular $SU(3)$ and $SO(3)$ flavour models. 

\begin{figure}[h]
\centering
\includegraphics[width=5.5cm]{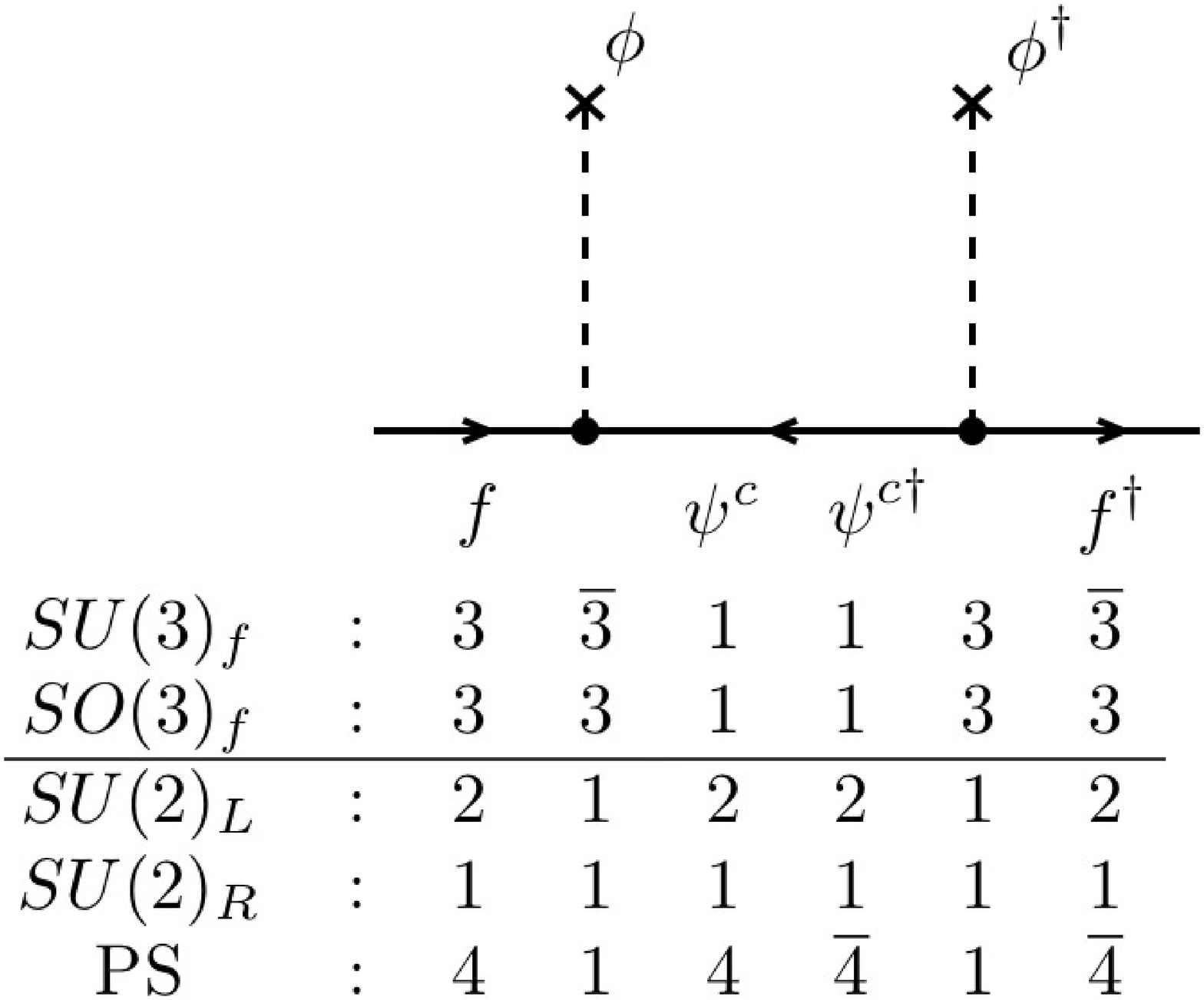} 
 \caption{A typical tree level correction to the K\"ahler of LH fermions generating the effective $k_{i}^{L}$ couplings driving the canonical normalisation corrections to the tri-bimaximal lepton mixing pattern.\label{Kahler}}
\end{figure}

Recall first that the $k_{0}^{L}$ coefficient governs the ``canonical'', i.e.\ renormalisable contribution in the K\"ahler $\propto \partial^{\mu}\tilde{L}^{\dagger}\partial_{\mu}\tilde{L}$ (for scalars) while $k_{3}^{L}$ emerges at higher order via operators like $\frac{1}{M_{K}^{2}}\partial^{\mu}\tilde{L}^{\dagger}\partial_{\mu}\tilde{L}\phi_{3}^{\dagger}\phi_{3}$ only and therefore is sensitive to the relevant messenger sector masses. Second, due to the self-conjugated structure of this type of operators, any messenger $\psi, \psi^{c}$ relevant for the Yukawa sector operators, i.e.\ with simultaneous couplings to flavon and matter superfields (like e.g.\ $\hat L \hat\phi \hat\psi^{c}$) necessarily enters the matter sector K\"ahler metric via effective operators of the form $\frac{1}{M^{2}_{\chi}}\partial^{\mu}\tilde{L}^{\dagger}\partial_{\mu}\tilde{L}\phi^{\dagger}\phi$ because no symmetry forbids such structures, c.f.~Figure \ref{Kahler}.

Since $SU(2)_{L}$ must remain intact upon flavour symmetry breaking, the messengers potentially affecting $k_{3}^{L}$ must necessarily be $SU(2)_{L}$-doublets, otherwise they can not couple to $\hat L \hat \phi$. In what follows, we shall namely check whether the $SU(2)_{L}$-doublet part of the messenger sector (if any) in the popular models can be naturally made heavy compared to the $SU(2)_{L}$-singlet messenger fields (transforming as $SU(2)_{R}$-doublet in the PS approach).

\begin{figure}[h]
\centering
case 1:\;\;\;\parbox{8cm}{\includegraphics[width=8cm]{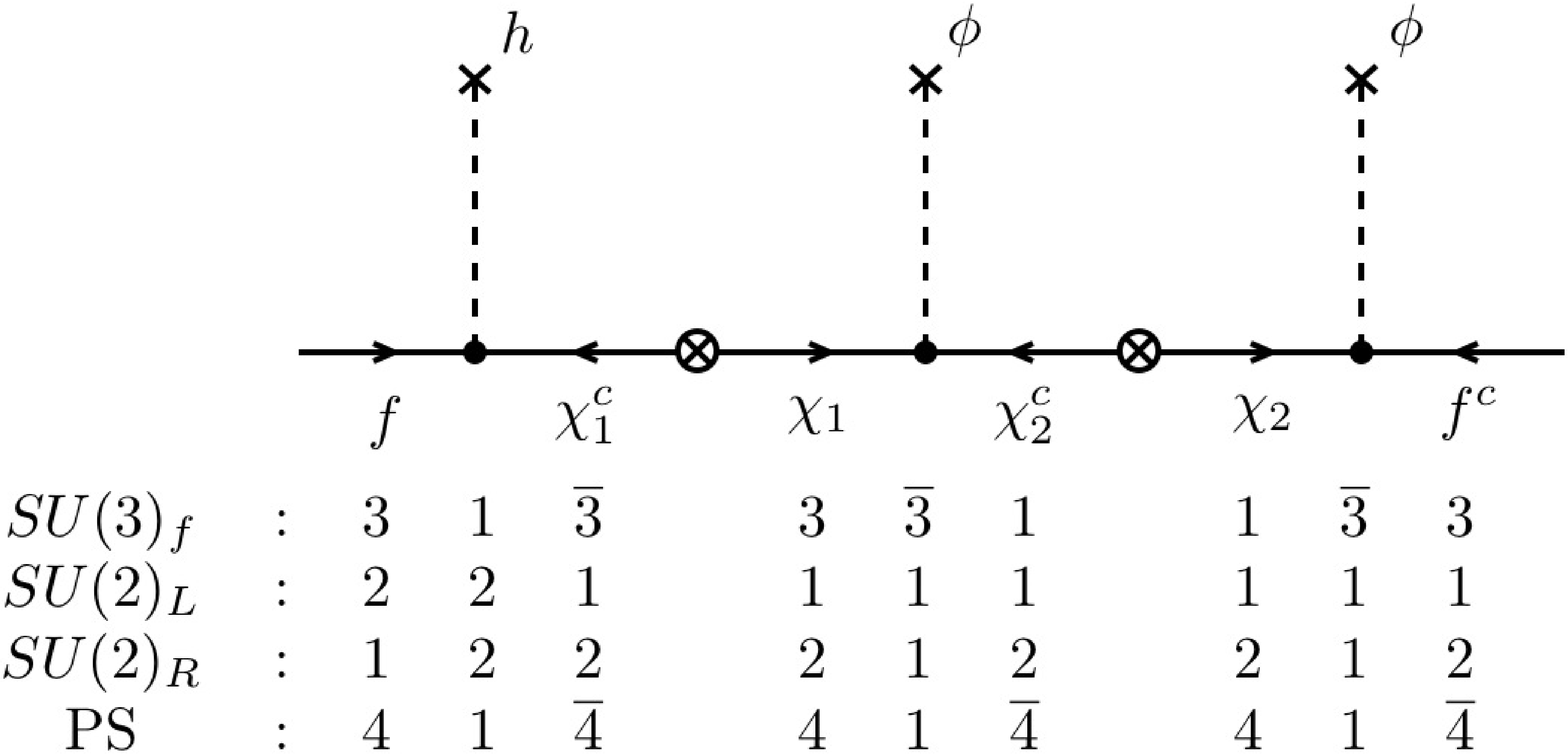}}\\ 
case 2:\;\;\;\parbox{8cm}{\includegraphics[width=8cm]{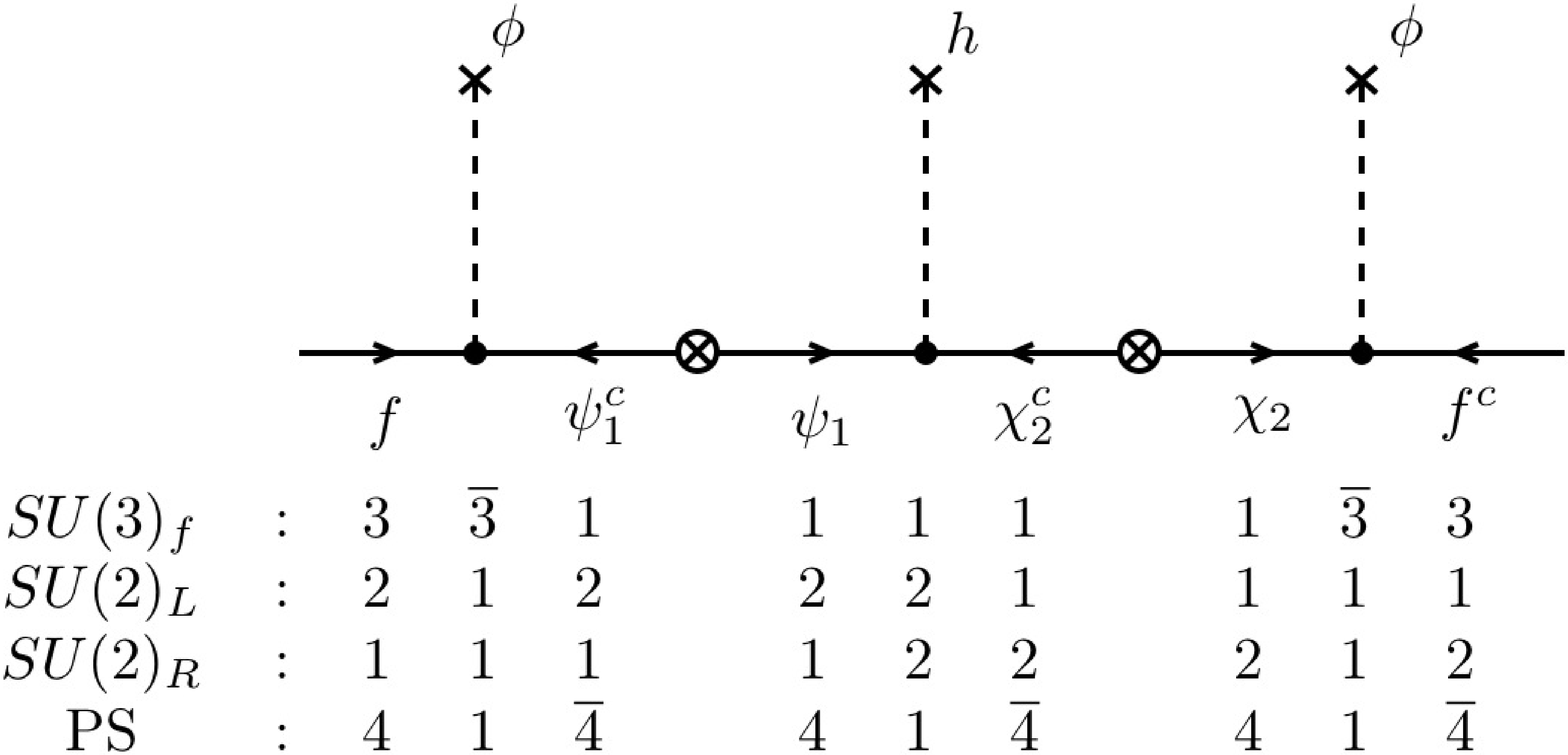}}\\ 
case 3:\;\;\;\parbox{8cm}{\includegraphics[width=8cm]{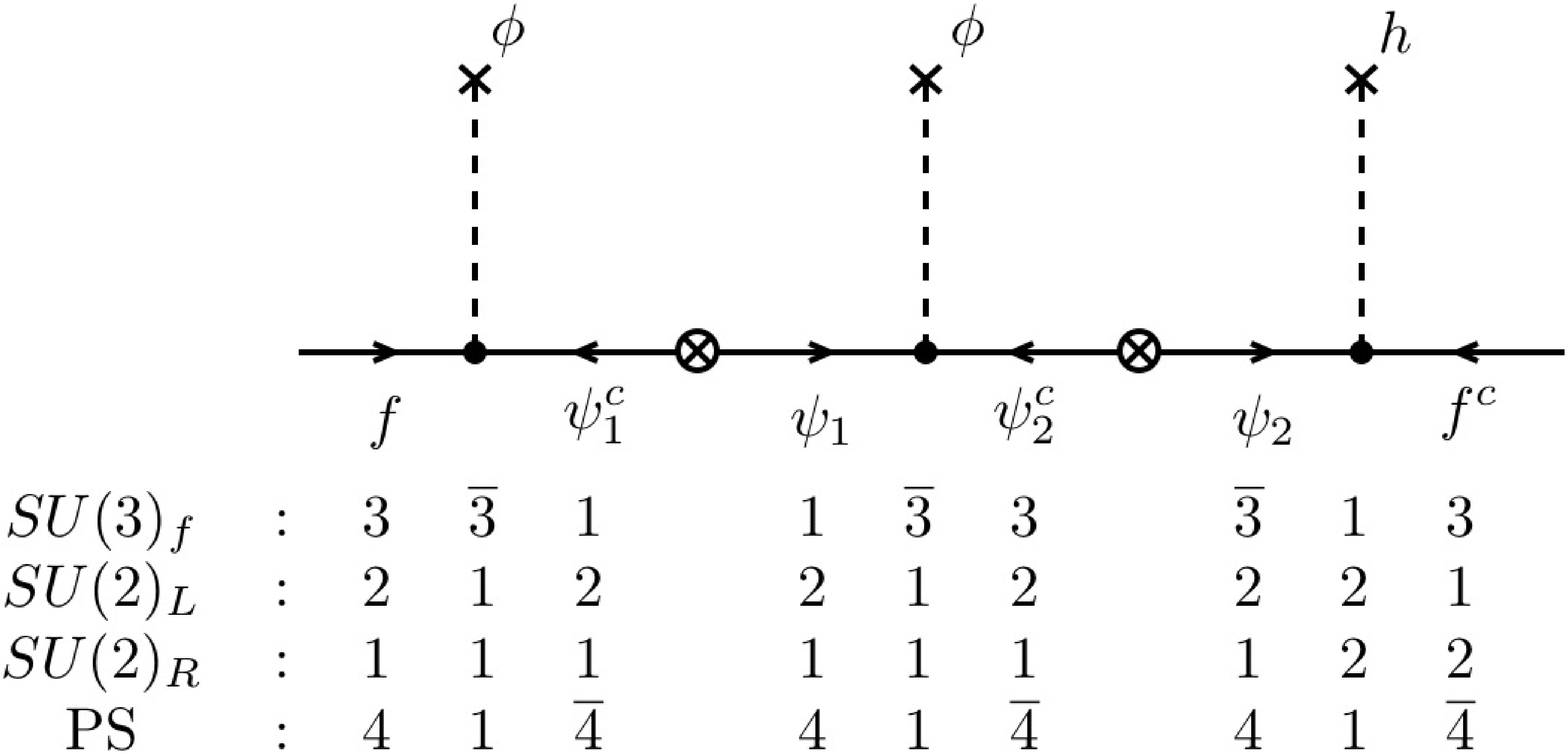}} 
 \caption{The three basic configurations of the messenger sector leading to the lowest order Yukawa sector effective operators in a typical $SU(3)$ flavour model. The position the SM Higgs VEV enters determines the $SU(2)_{L}\otimes SU(2)_{R}$ quantum numbers of the underlying messenger sector. We have used $\psi_{i}$ for the $SU(2)_{L}$ doublets while $\chi_{i}$ for the $SU(2)_{R}$ doublets respectively. \label{threestructures}}
\end{figure}

\subsection*{Models with $SU(3)$ family symmetry}
Starting with models based on $SU(3)$ family symmetry (or its discrete subgroups like $\Delta_{27}$), the triplet nature of both matter chiralities $f,f^{c}$ calls for a pair of antitriplet flavon insertions (up to the singular case of ${\bf 3}_{f} .{\bf 3}_{f^{c}}.{\bf 3}_{\phi}$-type contractions) so that the simplest Yukawa couplings have the internal structure depicted at Figure \ref{threestructures} (for discussions of the messenger sector of SU(3) models, see e.g.\ \cite{SU(3)messenger}).

The usual strategy in order to keep the  particle content of a model minimal is to exploit just some of these topologies for all the Yukawa sector entries. Typically, the first alternative is chosen, because in such a case the spectra of the $\chi$-type of messengers are sensitive to the large scale $SU(2)_{R}$ breaking providing for a bit more freedom in the Yukawa sector construction. This actually works rather well for all but the 33 Yukawa entries, that are preferred close to each other, at odds with the scaling properties of the other Yukawa entries (driven by expansion factors $\varepsilon^{2}$ or $\overline{\varepsilon}^{2}$ with $\varepsilon\sim 0.05$ and $\overline\varepsilon\sim 0.15$ for the up- and down-type sector respectively) and thus calling for extra contributions. 

\begin{figure}
\centering
$$\includegraphics[width=7.3cm]{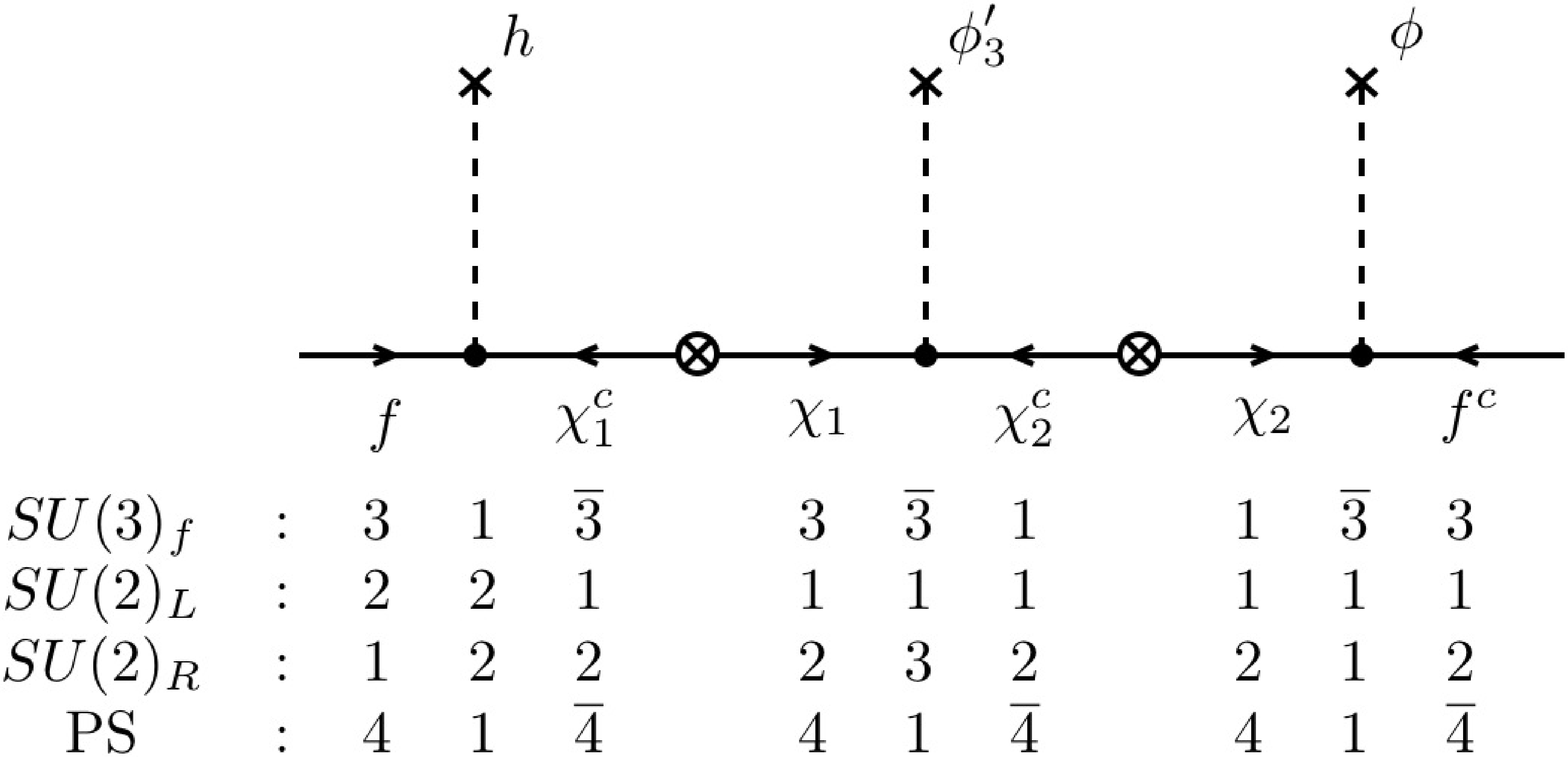} \quad
 \includegraphics[width=7.3cm]{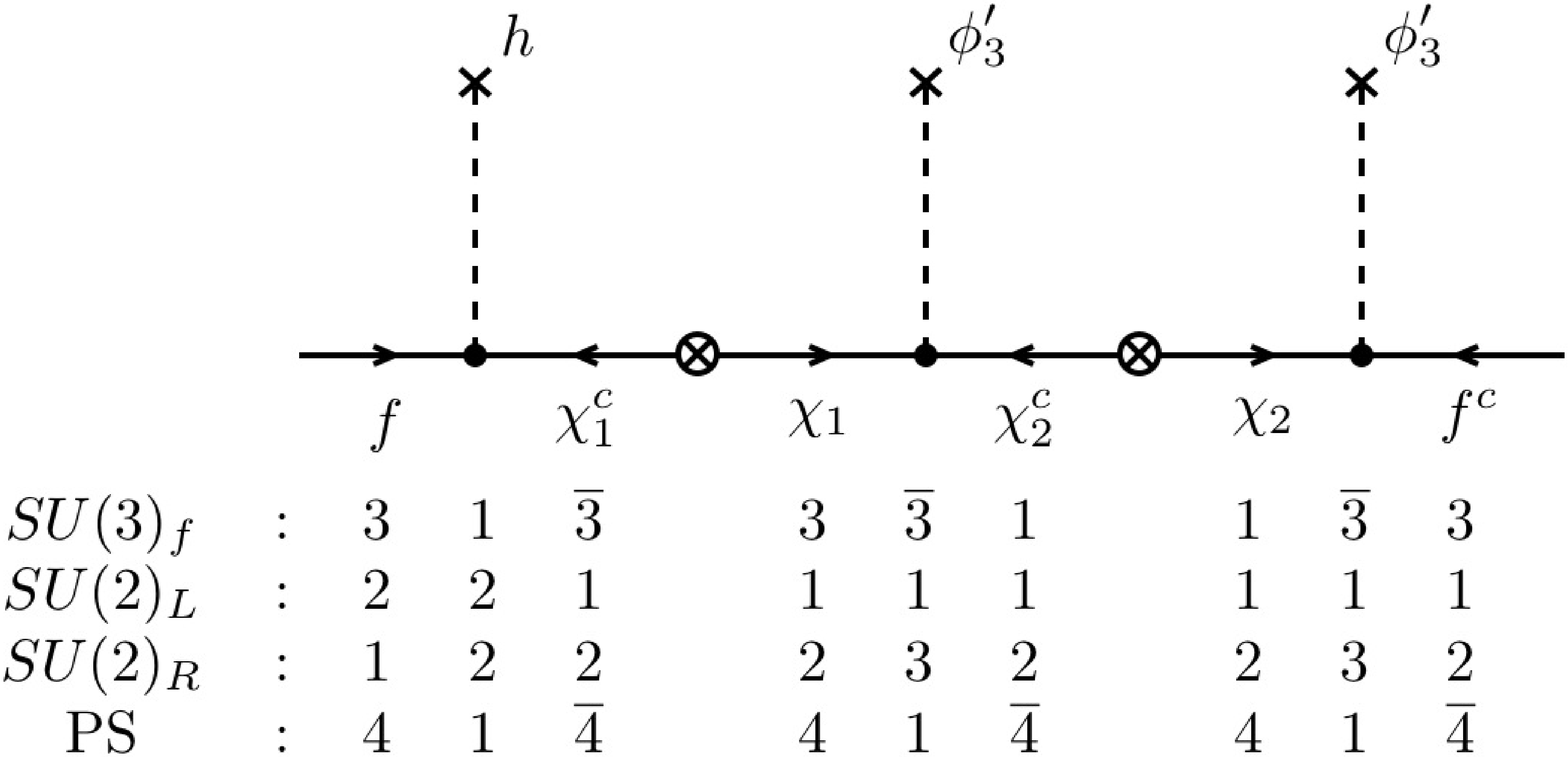} 
  $$
 \caption{Generating a pair of extra contributions to the 33 Yukawa entries by means of an extra $SU(2)_{R}$ triplet of $\phi_{3}$-type flavon fields $\phi_{3}'$.\label{tripletflavon} Notice that the messenger sector is $SU(2)_{L}$-singlet and thus does not trigger a potentially large violation of tri-bimaximal mixing in the lepton sector.}
\end{figure}

Such terms can then come from either an extra $\phi_{3}$-type flavon entering the graph of the same type (i.e.\ case 1 in Figure \ref{threestructures}) which has been exploited e.g.\ in the $SU(3)$ model by Varzielas-Ross \cite{deMedeirosVarzielas:2005ax} by means of the particular $SU(2)_{R}$-structure of $\phi_{3}=1\oplus 3$ (c.f. Figure \ref{tripletflavon}), 
or from a more complicated messenger sector 
with a left-handed ``$\psi$-type'' messengers admitting the other (case 2,3 in Figure \ref{threestructures}) contributions to the 33 Yukawa coupling. However, with the latter choice, a relatively light ``left-handed'' messenger must be postulated, leading to the instability of the tri-bimaximal lepton sector mixing generated by a potentially large deviation from universality in the $K_{L}$ part of the K\"ahler metric.

Thus, in order to avoid the potentially dangerous light $SU(2)_{L}$-doublet messengers $\psi$ one should base the effective Yukawa sector on the topologies of type 1 in Fig. \ref{threestructures}  that, however, comes for the price of extending the third-family flavon sector along the lines of \cite{deMedeirosVarzielas:2005ax} and further complication in the vacuum alignment mechanism. 

\subsection*{Models with $SO(3)$ family symmetry} 
\begin{figure}
\centering
case 1:\parbox{6cm}{\includegraphics[width=6cm]{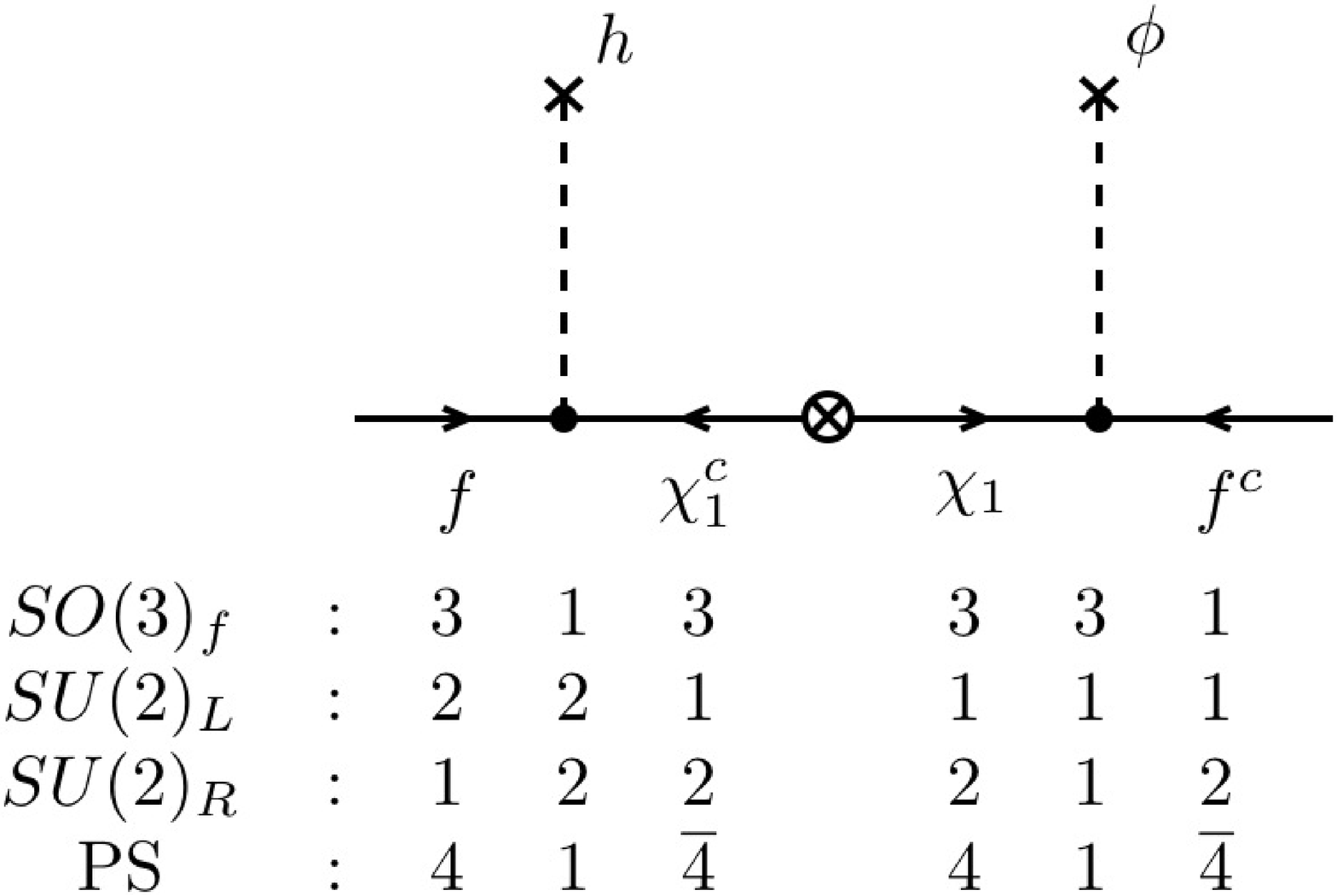}}\;\;\;\;\;\;
case 2:\parbox{6cm}{\includegraphics[width=6cm]{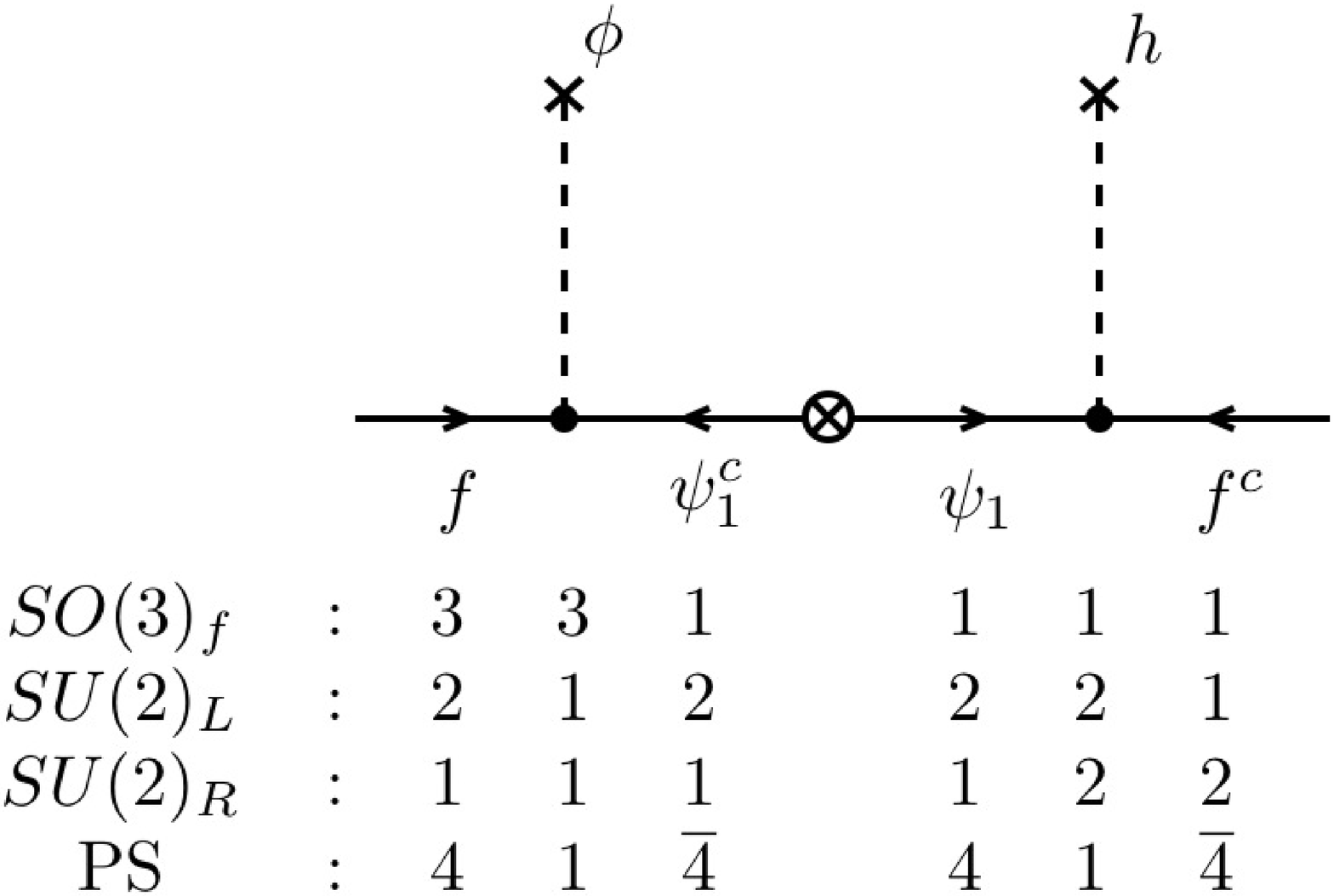}}\\ 
 \caption{The two basic configurations of the messenger sector leading to the lowest order Yukawa sector effective operators in a typical $SU(3)$ flavour model. The position of the SM Higgs VEV determines the $SU(2)_{L}\otimes SU(2)_{R}$ quantum numbers of the underlying messenger fields. As before, we have used $\psi_{i}$ for the $SU(2)_{L}$ doublets while $\chi_{i}$ for the $SU(2)_{R}$ doublets respectively. Note that the $\psi_{1}$ messenger in case 2 is usually ``flavon-specific'' and it is possible to forbid all the unwanted $\phi_{123}$, $\phi_{23}$ type of insertions by just a proper choice of the messenger sector quantum numbers.\label{SO3generic}}
\end{figure}

The situation in models based on $SO(3)$ is slightly simplified by the fact that the basic nontrivial singlet structure can be built out of two rather than three triplets without complex conjugation. Thus, in order to get realistic Yukawa patterns, only one chiral component (typically $f$) should transform as a triplet while the other as an $SO(3)$-singlet \cite{King:2006np,Antusch:2004xd,King:2005bj}\footnote{This, however, leads to problems with universality of the right-handed soft masses in SUSY, that in potentially realistic setups must be addressed by further assumptions.}. At the lowest level (in number of flavon insertions), we are left with only two basic options depicted in Fig.~\ref{SO3generic}.

\begin{figure}
\centering
\parbox{6cm}{\includegraphics[width=6cm]{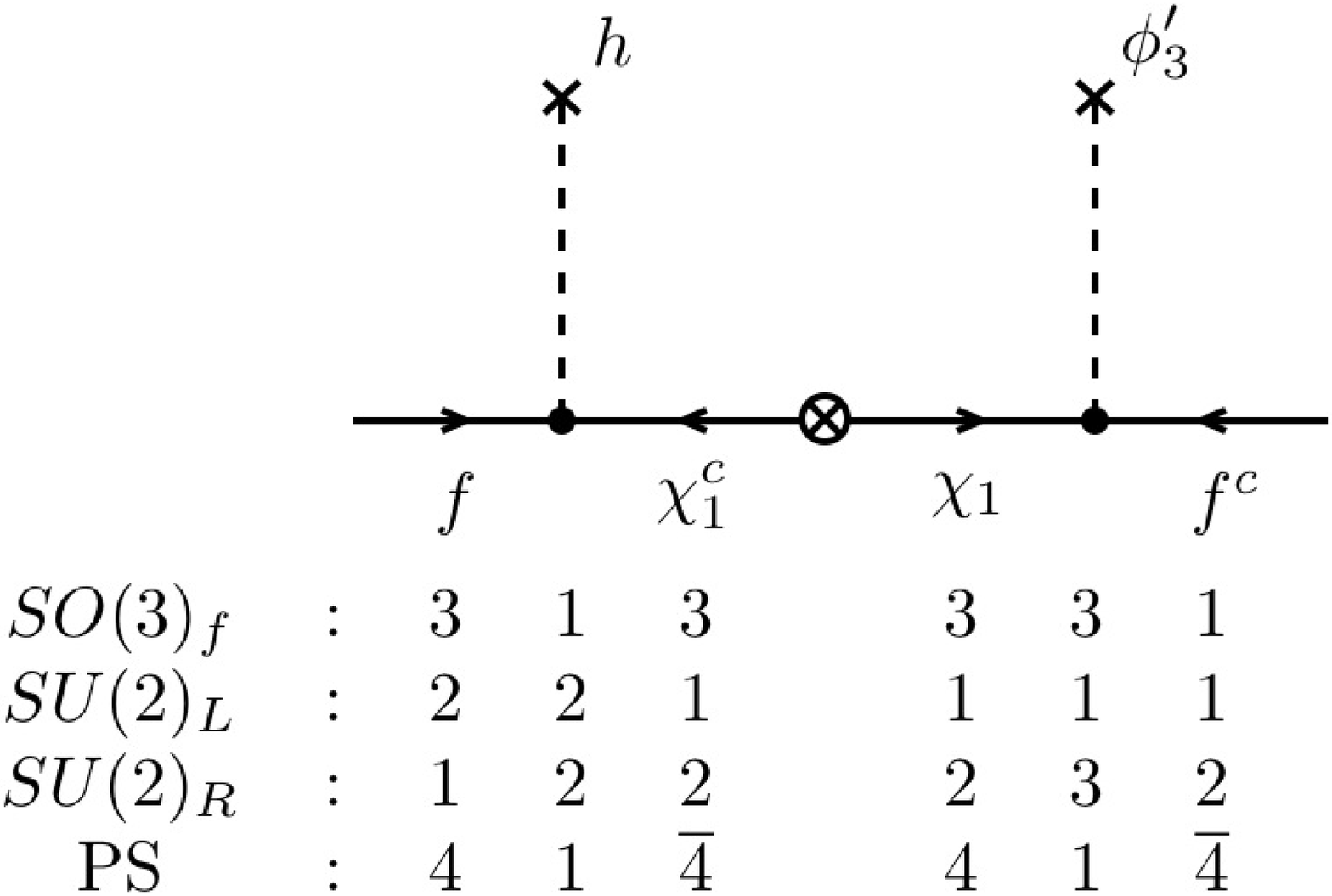}}\\ 
 \caption{Extra $\phi_{3}$-type flavon $\phi_{3}'$ as a solution to the Yukawa 33 entry irregularity without the unwanted LH K\"ahler correction effects in an $SO(3)$ flavour model.\label{SO3extraflavon}}
\end{figure}

Again, one can utilise the right-handed messengers (i.e.\ doublets of $SU(2)_{R}$) to obtain most of the desired Yukawa structures,  however, the above mentioned ``irregularity'' in the 33 entries calls for an extra contribution as in the $SU(3)$ case. Again, the basic options are either adding a left-handed (i.e.\ $SU(2)_{L}$ doublet ) messenger sector fields along case 2 indicated at Figure \ref{SO3generic}, c.f. \cite{King:2006np}, with potential impact on the left-handed K\"ahler corrections, or employ an extra $\phi_{3}$-type flavon, c.f. Figure \ref{SO3extraflavon}.

To conclude, the popular $SU(3)_{f}$-based flavour models \`a la Ross and Varzielas \cite{deMedeirosVarzielas:2005ax,deMedeirosVarzielas:2006fc} do not in general suffer from large K\"ahler corrections to the lepton sector tri-bimaximal mixing pattern due to the mere absence of the potentially dangerous $SU(2)_{L}$ doublets in the messenger sector. On the other hand, the $SO(3)$-class of models in versions \cite{King:2006np} can lead to substantial K\"ahler corrections because of employing a relatively light $SU(2)_{L}$ doublet in the messenger sector to resolve the 33 Yukawa issue. However, these models can be cured easily by invoking instead the ``extra $\phi_{3}$-type flavon solution'' with only $SU(2)_{R}$-doublet light messengers entering the K\"ahler metric along the lines sketched above.

%=======================================================  
\section{Comparison to other corrections to fermion mixings}\label{other}
%=======================================================
In this section, we would like to set the CN corrections in context to other corrections to fermion mixing, focusing on the corrections in the lepton sector. We will first consider effects from renormalisation group (RG) running and then from charged lepton mixing contributions. 

%=======================================================
\subsection{Renormalisation group corrections to tri-bimaximal neutrino mixing\label{RGsection}}
%=======================================================
The predictions for the Yukawa matrices arise at the scale of flavour symmetry breaking $\Lambda_F$, 
which we will assume to be close to the GUT scale ($M_{\mathrm{GUT}}$). In order to test such predictions experimentally, the renormalisation 
group (RG) running between $\Lambda_F$ and the electroweak scale $M_{Z}$ has to be taken into account. 
In particular, if tri-bimaximal mixing is realised in the neutrino sector, deviations from this pattern are
induced by RG running. 
The accurate calculation of such corrections requires evolving the 
effective neutrino mass matrix from $\Lambda_F \approx M_{\mathrm{GUT}}$ to low 
energy using the $\beta$-functions for the energy 
 ranges above and between the see-saw scales 
 and below the mass scale of the lightest right-handed neutrino   
\cite{RGE,RGEanalytical,Dighe:2006sr}. Numerically, this can be
done conveniently using the software package REAP \cite{Antusch:2005gp}. 

In what follows, we shall be
interested mainly in estimating the size of the RG corrections in the case of a hierarchical neutrino 
spectrum in the MSSM, for which the running effects are comparatively small and where the leading logarithmic 
approximation works reasonably well. Note that  due to the non-renormalisation theorem, only the radiative wave-function corrections contribute to the $\beta$-functions in supersymmetric theories and
the RG corrections can be treated in a very similar fashion to the canonical normalisation corrections, as we
will see explicitly below.

Following the spirit of section 4, let us consider the case when the wave-function renormalisation due to the 3rd
family dominates. More explicitly, we will assume that the 33-elements govern both $Y^e$ and  
$Y^\nu$ in the model basis (with diagonal $M_{M}$). This is the case, for instance, in the classes of non-Abelian flavour
models discussed in \cite{Antusch:2007re} (and in the example given in section \ref{effectsonsumrules}). Therefore, we will 
take:
$\label{Eq:YeYnuApprox33}
Y^e \approx \mbox{diag}(0,0,y_\tau)$ and $Y^\nu\approx \mbox{diag}(0,0,y_{\nu_{3}})
$.
Above the mass threshold of the heaviest RH neutrino $M_3$, the $\beta$-function for the 
effective neutrino mass matrix $m_\nu(\mu) = - v_u Y^\nu(\nu) M_{M}^{-1}(\mu) Y^{\nu T}(\mu)$ (where $\mu$
is the renormalisation scale and $v_u$ is the VEV of the up-type Higgs doublet) reads:
\begin{eqnarray}
16 \pi^2 \mu \frac{d}{d \mu} m_\nu &=& \left( Y^{e} Y^{e\dagger} + Y^{\nu} Y^{\nu \dagger}\right) m_\nu +
m_\nu \left( Y^{e*} Y^{eT} + Y^{\nu*} Y^{\nu T}  \right) \nonumber \\
& + & \left(-\frac{6}{5}g_{1}^{2}-6g_{2}^{2}+2\Tr Y^{\nu\dagger}Y^{\nu}+6\Tr Y^{u\dagger}Y^{u}\right) m_\nu\label{RGE} \;,
\end{eqnarray}
where the last term is proportional to the unit matrix in flavour space. Below $M_3$, the same $\beta$-function applies with $Y^\nu=0$.

Keeping at leading order all terms (but $m_{\nu}$) on the RHS of (\ref{RGE}) constant, one can integrate (\ref{RGE}) analytically, yielding
\bea
m_{\nu}(M_{Z})& =& m_{\nu}(M_{\mathrm{GUT}}) -m_{\nu}(M_{\mathrm{GUT}})\frac{1}{16\pi^{2}}\left[2\Tr Y^{\nu\dagger}Y^{\nu} \ln \frac{M_{\mathrm{GUT}}}{M_3} \right.\nn\\
&+ &
\left.\left(-\frac{6}{5}g_{1}^{2}-6g_{2}^{2}+6\Tr Y^{u\dagger}Y^{u}\right) \ln  \frac{M_{\mathrm{GUT}}}{M_{Z}} \right]\nn\\
& - & m_{\nu}(M_{\mathrm{GUT}})\frac{1}{16 \pi^2} \left(Y^{e*} Y^{eT} \ln \frac{M_{\mathrm{GUT}}}{M_{Z}}
+ Y^{\nu*} Y^{\nu T} \ln  \frac{M_{\mathrm{GUT}}}{M_3} \right)\nn\\
& - & \frac{1}{16 \pi^2} \left(Y^{e\dagger} Y^e \ln  \frac{M_{\mathrm{GUT}}}{M_{Z}} 
+ Y^{\nu\dagger} Y^\nu \ln \frac{M_{\mathrm{GUT}}}{M_3}\right)m_{\nu}(M_{\mathrm{GUT}})\nn+\ldots,
\eea
which can be rewritten as (forgetting about the doubly-suppressed mixed terms):
\begin{eqnarray}\label{Eq:RGEfor33dominanceLLog}
m_\nu(M_{Z}) \approx P^{T}_{\mathrm{RG}} \: m_\nu({M_{\mathrm{GUT}}}) \: P_{\mathrm{RG}}\qquad \mathrm{with} \qquad P_{\mathrm{RG}}=r \mathbbm{1} + \Delta P_\mathrm{RG}+\ldots\;,
\end{eqnarray}
where
\begin{eqnarray}
r &=& 1-\frac{1}{16\pi^{2}}\left[\Tr Y^{\nu\dagger}Y^{\nu} \ln \frac{M_{\mathrm{GUT}}}{M_3}+3\left(-\frac{1}{5}g_{1}^{2}-g_{2}^{2}+\Tr Y^{u\dagger}Y^{u}\right) \ln \frac{M_{\mathrm{GUT}}}{M_{Z}}\right],\nn\\
\Delta P_{\mathrm{RG}} &= & 
-  \frac{1}{16 \pi^2}\left[Y^{e*} Y^{eT} \ln \frac{M_{\mathrm{GUT}}}{M_{Z}}
+ Y^{\nu*} Y^{\nu T} \ln  \frac{M_{\mathrm{GUT}}}{M_3} \right]  \;.
\end{eqnarray}
Note that the $r$-factor in (\ref{Eq:RGEfor33dominanceLLog}) is irrelevant for the lepton mixing, because at the leading order one can rewrite $P_{\mathrm{RG}}$ in the form
\be
 P_{\mathrm{RG}}=r\left(\mathbbm{1}+\Delta P_{\mathrm{RG}}\right)+{\cal O}\left[\frac{1}{(16\pi^{2})^{2}}\right]\;\mathrm{terms},
\ee
but overall factors like $r$ drop in formula (\ref{DeltaWnu}). 

As we mentioned, the leading order RG effect (\ref{Eq:RGEfor33dominanceLLog}) has exactly the form of Eq.~(\ref{cannormmatrices}), so both types of corrections, from RG running, as well as  from canonical normalisation, can be treated on the same
footing in this approximation. Furthermore, using Eq.~(\ref{Eq:YeYnuApprox33}) and comparing 
Eq.~(\ref{Eq:RGEfor33dominanceLLog}) with Eq.~(\ref{DeltaP}), we find that there is again a single parameter governing
the RG corrections to all the mixing angles given by:
\begin{eqnarray}\label{Eq:etaRG}
 \eta^{\mathrm{RG}} =
\frac{y^2_\tau}{8 \pi^2}\ln \frac{M_{\mathrm{GUT}}}{M_{Z}}  
+ \frac{y_{\nu_{3}}^2}{8 \pi^2} \ln \frac{M_{\mathrm{GUT}}}{M_3}\;.
\end{eqnarray}
The quantitative predictions of the RG running effects can then be obtained from the relevant formulae for the CN corrections (\ref{subleading}), (\ref{correctionTBcase}), (\ref{rsa}), (\ref{13sumruleprimed}), (\ref{sumrulefinished}), upon swapping $\eta^\mathrm{K} \leftrightarrow\eta^{\mathrm{RG}}$. The last contribution (\ref{Eq:etaRG}) would be absent if $M_3 > M_{\mathrm{GUT}}$. 

We have cross-checked these results with the analytic
approximations presented in \cite{RGEanalytical} and found a perfect
agreement for the considered case. 
In summary, with tri-bimaximal neutrino mixing at the GUT scale, the low scale parameters are given
approximately by:
\be
s^\nu_{12}({M_{Z}}) = \frac{1}{\sqrt{3}}\!\!\left( 1 + \frac{\eta^{\mathrm{RG}}}{6}\right), \;
s^\nu_{23}({M_{Z}}) = \frac{1}{\sqrt{2}}\!\!\left( 1 + \frac{\eta^{\mathrm{RG}}}{4}\right), \;
s^\nu_{13}({M_{Z}}) \propto \eta^{\mathrm{RG}} \frac{m_2}{m_3}\;.
\ee

\subsection{Combined treatment of RG and canonical normalisation corrections}
Finally, one can even subsume the effects of 3rd family dominated RG and CN corrections to tri-bimaximal
mixing into a single physical parameter: 
\begin{eqnarray}
\label{etatotal}
\eta = \eta^{\mathrm{RG}} + \eta^{\mathrm{K}}\;,
\end{eqnarray}
where $\eta^{\mathrm{RG}}$ is defined in (\ref{Eq:etaRG}) and $\eta^{\mathrm{K}}$ is given in section \ref{Kahlercorrections}, c.f.\ Eq.~(\ref{Kahler33}). 
In the following, we will apply this combined treatment to discuss CN and RG effects in the presence of charged lepton mixing corrections to tri-bimaximal mixing.  Note that while the size of the RG effects depends mainly on $\tan \beta$ (which governs the size of $y_\tau$) and on the $M_3$ - $M_{\mathrm{GUT}}$ hierarchy, the size of the canonical normalisation corrections depends on the messenger sector as discussed in section \ref{models}.

%============================= ==========================
\subsection{Charged lepton mixing corrections to tri-bimaximal neutrino mixing}
%=======================================================
Assume that (in the basis in which $\hat V_{L}^{\nu\dagger}=U_{TB}$) there is a finite contribution to the lepton mixing matrix coming from the charged lepton sector, as it is actually common to many potentially realistic models of flavour employing unified gauge symmetries like Pati-Salam \cite{PS} or $SO(10)$ \cite{SO10}. The charged lepton sector mixing in such cases tends to copy the structure of $\hat V_{L}^{d}$ (up to Clebsch factors) that leads to a natural assumption about the structure of the $\hat V_{L}^{l}$ matrix (before the effects of canonical normalisation are taken into account):
\be\label{VLl}
\hat V_{L}^{l}\approx \begin{pmatrix}
\hat c_{12}^l & \hat s_{12}^le^{-i\hat \rho} & 0 \\
-\hat s_{12}^le^{i\hat \rho} & \hat c_{12}^l & 0 \\
0 & 0 & 1
\end{pmatrix},
\ee
where $\hat s_{12}^l$ is a small Cabibbo-like mixing (typically $\hat s_{12}^l\approx \lambda/3$) and $\hat \rho$ is a generic phase.
In such a case, the exact tri-bimaximal structure of the (high-scale) lepton mixing matrix is lifted and one is left  with (assuming as before $\hat V_{L}^{\nu\dagger}=U_{TB}$): 
\be\hat U_{PMNS}=\hat V_{L}^{l}\hat V_{L}^{\nu\dagger}=\hat V_{L}^{l}U_{TB}\;,
\ee 
up to a rephasing to the standard PDG form \cite{Yao:2006px}, which is needed due to the extra phase in (\ref{VLl}).
The charged lepton sector contribution (\ref{VLl}) has multiple effects, in particular breaks the direct link between the ``measured'' (up to the renormalisation group running \cite{RGEanalytical}) values of the lepton mixing parameters and the underlying purely neutrino sector rotations. However, due to the particular structure of $\hat V_{L}^{l}$ above (leading to just a mild alteration of the tri-bimaximal neutrino mixing pattern), a set of simple relations between the underlying neutrino and charged lepton sector mixings and $\hat U_{PMNS}$ can be obtained. 

In particular, the original zero reactor angle is lifted by the 12 rotation in the charged lepton sector to: 
\be\label{reactoranglelifting}
(\hat U_{PMNS})_{13}=- \hat s_{23}^{\nu}\hat s_{12}^{l}e^{-i\hat\rho} \quad \Rightarrow \quad \hat \theta_{13}\approx \hat s_{23}^{\nu}\hat s_{12}^{l}=\frac{1}{\sqrt{2}} \hat s_{12}^{l}
\ee
(no phases enter because we are looking at a magnitude of the 13 term only), which in Georgi-Jarlskog type of unified models \cite{Georgi:1979df} (where $\hat s_{12}^l = \theta_{C}/3$ with $\theta_{C}$ denoting the quark-sector Cabibbo mixing) yields:
\be\label{GJmodels}
\hat \theta_{13}\approx \frac{\theta_{C}}{3\sqrt{2}} \;.
\ee
Second, there is an interesting phenomenologically testable sum-rule for the deviation of the solar angle from its exactly tri-bimaximal value $\theta_{TB}=35^{o}16'$ in the form \cite{sumrule}
\be\label{mainsumrule}
\hat\theta_{12}=\theta_{12}^{TB}+\hat\theta_{13}\cos\hat\delta,\quad \mathrm{i.e.}\quad \hat s= \hat r \cos\hat\delta\;,
\ee 
where $\hat \delta$ stands for the Dirac CP phase in the lepton sector\footnote{This sum-rule can be easily derived (c.f. Appendix  \ref{sumrulederivation}) from the {\it magnitude} of the 31 entry of $\hat U_{PMNS}=\hat V_{L}^{l}U_{TB}$ and thus is insensitive to the Majorana phases.}. An interested reader can find the derivation of formulae (\ref{reactoranglelifting}) and (\ref{mainsumrule}) in Appendix \ref{sumrulederivation}.

%----------------------------------------------------------------------------------------
\subsection{Canonical normalisation corrections to lepton mixing sum-rules\label{effectsonsumrules}}
%----------------------------------------------------------------------------------------
In view of results of section \ref{simplestsetup}, let us discuss the stability of these formulae with respect to the effects of canonical normalisation. We shall again assume the (leading order) 33-sector non-universality in the corresponding K\"ahler metric (\ref{Kahler33}). Remarkably enough, though $\hat V_{L}^{l}$ is nontrivial, the block-structure of $K_{f}$ is such that $\hat V_{L}^{l}$ plays essentially no role in the leading-order formula (\ref{DeltaWl}) and one recovers (\ref{Wsimplest}) as in the simplest case discussed in the previous section. The canonical normalisation corrections to the lepton mixing matrix then obey: 
\be
\Delta U_{PMNS}=-i\hat U_{PMNS} \Delta W_{L}^{\nu\dagger},
\ee
where $\hat U_{PMNS}$ is not equal to $U_{TB}$ as in the simplest case, but  $\hat U_{PMNS}=\hat V_{L}^{l}U_{TB}$.
%-----------------------------------------------------------------------
\paragraph{Corrections to the charged-lepton-sector-induced 13 mixing:}\mbox{}\\
%-----------------------------------------------------------------------
Let us look first at the CN corrections induced in the simpler formula (\ref{reactoranglelifting}).
There is no a-priori reason the 13 entry of $U_{PMNS}$ should vanish as it was the case at the leading order in the purely tri-bimaximal setting (\ref{correctionTBcase}). Indeed, we have:
\be\label{firstturn}
\Delta (U_{PMNS})_{13}=-i(\hat U_{PMNS})_{11} (\Delta W_{L}^{\nu\dagger})_{13}-i(\hat U_{PMNS})_{12} (\Delta W_{L}^{\nu\dagger})_{23}\;,
\ee
giving at the leading order:
\bea
-i\sum_{i=1,2}(\hat U_{PMNS})_{1i} (\Delta W_{L}^{\nu})_{i3}&=&\sum_{i}\sum_{j}(\hat V_{L}^{l})_{1j}(U_{TB})_{ji}\left(U_{TB}^{\dagger}\Delta P_{L}^{T}U_{TB}\right)_{i3}\nn\\
& & -\sum_{j}(\hat V_{L}^{l})_{1j}(U_{TB})_{j3}\left(U_{TB}^{\dagger}\Delta P_{L}^{T}U_{TB}\right)_{33}\label{jsum}\;.
\eea
Due to unitarity and the shape (\ref{DeltaP}) of $\Delta P_{L}$, the first term on the RHS of (\ref{jsum}) is zero, while the latter yields:
\be\label{res}
\Delta (U_{PMNS})_{13}\approx(\hat V_{L}^{l})_{12}(U_{TB})_{23}\frac{\eta^\mathrm{K}}{4}=-\hat s_{12}^{l}e^{-i\hat \rho}\frac{\eta^\mathrm{K}}{4\sqrt{2}}\;.
\ee
Notice that the Majorana phase structure of this correction is again the same like the phase structure of the defining basis 13 entry\footnote{i.e.\ zero phase in the given global phase convention fixing the shape of $P_{M}$ with a real 33 entry.} in (\ref{reactoranglelifting}) and thus the Dirac CP phase is stable under CN effects. 
Taking into account also the subleading correction 
$\propto \sqrt{{\Delta m^{2}_{\odot}}/{\Delta m^{2}_{A}}}$ (which is of the same order as the term in Eq.~(\ref{res}))
of the type (\ref{subleading}), the last formula is extended to:
\be\label{res2}
\Delta (U_{PMNS})_{13}\approx 
-\frac{\eta^\mathrm{K}}{4\sqrt{2}}\left(e^{-i\hat\rho}\hat s_{12}^{l}+\hat c_{12}^{l}\frac{4}{3}e^{i\tfrac{\hat \alpha_{2}}{2}}\sqrt{\frac{\Delta m^{2}_{\odot}}{\Delta m^{2}_{A}}}\right),
\ee
where, as before, the $\alpha_{2}$ phase accounts for the extra phase ambiguity due to the Majorana nature of the neutrinos, c.f. discussion of formula (\ref{subleading}).
In order to deduce the K\"ahler correction to the `induced' 13 mixing (\ref{reactoranglelifting}), this result should be added to the RHS of formula (\ref{reactoranglelifting}) leading to:
\be\label{13sumruleprimed}
s_{13}\approx \left|\hat s_{12}^l \frac{1}{\sqrt{2}}\left(1+\frac{\eta^\mathrm{K}}{4}\right)-\hat c_{12}^l\frac{\eta^\mathrm{K}}{3\sqrt{2}}e^{i\left(\tfrac{\hat \alpha_{2}}{2}+\hat\rho\right)}\sqrt{\frac{\Delta m^{2}_{\odot}}{\Delta m^{2}_{A}}}\right|
\ee
for the physical $13$ mixing in $U_{PMNS}$. Notice that there is a slight ambiguity due to the phase factor in the second term, that can not be neglected with respect to the $\eta^\mathrm{K}$-part of the first term therein.
However, the smallness of the `charged-lepton-sector-induced' reactor mixing angle $s_{13}\approx \tfrac{1}{\sqrt{2}}\hat s_{12}^{l}$ (which is typically $\frac{\theta_{C}}{3\sqrt{2}}$ for Georgi-Jarlskog type of  flavour models corresponding to the first term above, c.f. formula (\ref{GJmodels}) and the discussion around) is not disturbed by the effects of canonical normalisation.
%=============================================================
\paragraph{Canonical normalisation corrections to $\hat s = \hat r \cos\hat\delta$ :}\mbox{}\\
%=============================================================
With the information about the $\theta_{13}$ stability at hand, one can infer the leading additive correction to the defining basis formula (\ref{mainsumrule}), that (swapping all the defining basis quantities for their physical counterparts) should read: 
\be\label{sumruleguess}
\theta_{12}^{TB}=\theta_{12}-\theta_{13}\cos\delta+f(\eta^\mathrm{K})\;,
\ee
where $f(\eta^\mathrm{K})$ is a linear function of $\eta^\mathrm{K}$ vanishing for $\eta^\mathrm{K}\to 0$, i.e.\ $f(\eta^\mathrm{K})=c\,\eta^\mathrm{K}$ with a real proportionality factor $c$. As we have seen in the previous paragraph, the leading CN correction to the 13-mixing (\ref{13sumruleprimed}) is only multiplicative (\ref{13sumruleprimed}) and thus all the would-be corrections in (\ref{sumruleguess}) due to the  $\eta^\mathrm{K}$-sensitivity in $\theta_{13}$ or $\cos\delta$ are suppressed by $\theta_{13}$. This means that in the $\theta_{13}\to 0$ limit in (\ref{sumruleguess}), one should recover the simple leading order $\theta_{12}$ scaling obtained  in section \ref{simplestsetup}. Thus, one  gets $c=-\frac{1}{6\sqrt{2}}$, which gives at the leading order:
\be\label{sumrulefinished}
\theta_{12}^{TB}=\theta_{12}-\theta_{13}\cos\delta-\frac{\eta^\mathrm{K}}{6 \sqrt{2}}\;.
\ee
Formula (\ref{sumrulefinished}) can be finally recast (using $a=\frac{\eta^\mathrm{K}}{4}$ derived in section \ref{simplestsetup}) into a sum-rule for {\it measurable quantities} $a$, $s$ and $r$ and $\delta$ only:\footnote{
We note that the sum rule of Eq.~(\ref{mainresult}) can be readily generalised to arbitrary $\hat\theta^\nu_{12}$ (but keeping $\hat\theta^\nu_{13}=0^\circ$ and $\hat\theta^\nu_{23}=45^\circ$ fixed) using $s^\nu_{12}=\hat s^\nu_{12}(1+\eta^K (\hat c^\nu_{12})^2 /4)$ and repeating the derivation of Appendix \ref{sumrulederivation}.   
}
\be\label{mainresult}
s=r\cos\delta+\frac{2}{3}a\;.
\ee
This relation identifies a characteristic imprint of the canonical normalisation effects in the popular scheme where the charged leptons contribute in the Georgi-Jarlskog manner (i.e.\ only the 12 sector rotation is non-negligible) while the neutrino sector mixing is exactly tri-bimaximal. 
Note that in addition to the precision measurements required for testing the original sum rule  
\cite{sumrule}, testing equation~(\ref{mainresult}) requires an accurate measurement of the deviation from maximal atmospheric mixing \cite{Antusch:2004yx}.

The simple argument above can only be used to fix the shape of the leading order {\it additive} corrections in (\ref{sumruleguess}) and derive the main result (\ref{mainresult}), but does not, in general, provide any information on sub-leading corrections (entering either as multiplicative changes in small parameters or higher order effects in $\eta^\mathrm{K}$) to (\ref{sumruleguess}).
An interested reader can find a more detailed explicit derivation of (\ref{sumrulefinished}) with a brief discussion of the shape of such subleading corrections in the Appendix.
Remarkably enough, this formula is {\it stable} also under radiative corrections due to the RG running (see in section \ref{RGsection}), which makes it directly testable at future experimental facilities.

%=========================================================================
\section{Summary and discussion}
%=========================================================================

In summary, we have re-analysed the effect of canonical normalisation
of kinetic terms on the quark and lepton mixing angles.
In contrast to previous studies we have found that the effects
can lead to significant corrections to the fermion mixing angles. 
Such potentially large effects
are characteristic of flavour models based on non-Abelian family
symmetries, where some of the possible K\"ahler potential (and
superpotential) operators, in particular those associated with the
third family, are only mildly suppressed. We have investigated
under which conditions the messenger sector of such flavour models
generates such K\"ahler potential operators for which canonical
normalisation effects are sizeable, and under which conditions these
operators may be absent and canonical normalisation effects are
small. The quantitative significance of the canonical normalisation
effects is clearly model dependent, and in order to address this we have
provided a detailed discussion of the messenger sectors responsible
for both the K\"ahler potential and the superpotential corrections in
the class of $SU(3)$ and $SO(3)$ flavour models. For example in 
the $SU(3)$ or $\Delta_{27}$ models
\cite{deMedeirosVarzielas:2005ax}, the left-handed messengers sector
essentially decouples from the effective Yukawa couplings and the
K\"ahler metric for the left-chirality matter fields is only subject
to small corrections leading to $\eta^{K}\sim 0$ and thus negligible
CN effects.  On the other hand, in the $SO(3)$ or $A_4$ models
\cite{King:2006np}, the left-handed messengers $\psi$ have been
assumed to be quite light, in which case the wave-function effects of
third family rescaling described in this paper are expected to be
large with $\eta^{K}\sim {\cal O}(1)$.

We developed a general perturbative
formalism which enables the CN effects in both quark and lepton
sectors to be estimated. We then applied this formalism to 
explicit examples for potentially relevant CN effects.
For example, we 
have discussed the corrections to the CKM matrix element $|V_{cb}|$ as
well as corrections to tri-bimaximal neutrino mixing.  
In the quark sector we found that such
canonical normalisation effects could imply a relatively large change in
$V_{cb}$, that (although still only multiplicative) could be much
larger than the estimates given previously in the literature where the
possibility of a large third family expansion parameter was not
considered. Concerning leptons, we found that the physical effect
of canonical normalisation is
strongly amplified compared to the quark sector, because of the approximate 
tri-bimaximality of the solar and atmospheric mixings. On the other hand, the
(comparatively) small reactor angle receives only sub-leading corrections.

We have also compared the CN corrections with other relevant corrections to 
predictions of flavour models.
Regarding renormalisation group (RG) corrections in leading logarithmic approximation, 
we have expressed the effects in a form which allows a perturbative treatment
analogous to the one used for our analytical estimates of the CN effects. We have shown how, in 
the case that third family effects dominate RG and CN corrections, 
both sorts of corrections can be subsumed into a single universal parameter at leading order. 
As application we have presented a detailed discussion of such third family effects 
on the lepton mixing sum rule $s = r \cos \delta$ \cite{sumrule} which emerges as a 
relation among lepton sector observables if the leading neutrino sector mixing is 
exactly tri-bimaximal and modified only by small (but relevant) Cabibbo-like 
charged lepton mixing contributions. In this sum rule $s,r$ describe the deviations 
of solar and reactor mixing angles from their tri-bimaximal values, and $\delta$ is 
the observable Dirac CP phase in PDG parameterisation \cite{Yao:2006px}. 
Assuming hierarchical neutrinos and taking into account 
both, CN and RG third family wave-function effects, we have
discussed in detail how the stable version of the sum rule 
$s = r \cos \delta + \frac{2}{3} a$ \cite{Antusch:2007ib} 
is derived (presenting additional details of the derivation, beyond the previous 
analysis). The additional parameter $a$ in the stable sum rule accounts for the 
deviation of the atmospheric mixing angle from its (tri-bi)maximal value $\pi/4$ 
due to the combined third family CN and RG effects.

In conclusion, the main message of this paper is that
in certain classes of models canonical normalisation effects,
in particular those associated with the third family,
may be larger than previously thought, leading to 
larger corrections to quark and lepton mixing angles
than previously realised.

\section*{Acknowledgements}
We acknowledge partial support from the following grants:
PPARC Rolling Grant PPA/G/S/ 2003/00096;
EU Network MRTN-CT-2004-503369;
EU ILIAS RII3-CT-2004-506222;
NATO grant PST.CLG.980066.

\appendix
\section*{Appendices}
\section{Conventions - CKM \& PMNS mixing matrices}
In general, the mixing matrix in the lepton sector, the PMNS matrix
$U_{\mathrm{PMNS}}$, is defined as the matrix appearing in the
charged electroweak currents expressed in terms of lepton
mass eigenstates. Denoting the charged lepton mass matrix by
$M_\mathrm{l}$ and the light neutrino mass matrix by $m_{\nu}$, the mass part of the matter sector lagrangian reads:
\begin{eqnarray}
{\cal L}=-  \bar{L}_L M_l l_R  
- \tfrac{1}{2}\bar{\nu}_L m_{\nu} \nu_\mathrm{L}^c 
+ \text{H.c.}\; .
\end{eqnarray}
Performing the transformation from flavour to mass basis by
 \begin{eqnarray}\label{eq:DiagMe}
V_{L}^{l} \, M_l \,
V^{l\dagger}_{R} =
\mbox{diag}(m_e,m_\mu,m_\tau)
 , \quad
V_{L}^{\nu} \,m_\nu\,V^{\nu T}_{L} =
\mbox{diag}(m_1,m_2,m_3),
\end{eqnarray}
the PMNS matrix is given by
\begin{eqnarray}\label{Eq:PMNS_Definition}
U_{{PMNS}} 
= V_{L}^{l} V_{L}^{\nu\dagger} \,.
\end{eqnarray}
Here it is assumed implicitly that unphysical phases are removed by
field redefinitions, and $U_\mathrm{PMNS}$ contains one Dirac phase
and two Majorana phases\footnote{The latter are physical only in the case of
Majorana neutrinos, for Dirac neutrinos the two Majorana phases can be
absorbed as well.}. 

The standard PDG parameterisation of the
PMNS matrix (see e.g.\cite{Yao:2006px}) is: 
\begin{eqnarray}\label{Eq:StandardParametrisation}
 %\hspace{-0.5cm} 
 U_{\mathrm{PMNS}} = \left(
  \begin{array}{ccc}
  c_{12}c_{13} & 
  s_{12}c_{13} & s_{13}e^{-i \delta}\\
  -c_{23}s_{12}-s_{13}s_{23}c_{12}e^{i \delta} &
  c_{23}c_{12}-s_{13}s_{23}s_{12}e^{i \delta}  &
  s_{23}c_{13}\\
  s_{23}s_{12}-s_{13}c_{23}c_{12}e^{i \delta} &
  -s_{23}c_{12}-s_{13}c_{23}s_{12}e^{i \delta} &
  c_{23}c_{13}
  \end{array}
  \right) \, P_M\, ,
\end{eqnarray}
which is used in most analyses of neutrino oscillation experiments.
Here $\delta$ is the Dirac CP violating phase which is in
principle measurable in neutrino oscillation experiments, and 
$P_{M} = \mathrm{diag}(e^{i  \tfrac{\alpha_1}{2}}, 
e^{i \tfrac{\alpha_2}{2}}, 1)$ contains the two measurable Majorana phase differences $\alpha_1,
\alpha_2$.  In the body of this manuscript we use this standard
parameterisation also for 
$V^{\nu\dagger}_{L}$ and denote the
corresponding mixing angles by $\theta_{ij}^\nu$, while the mixing
angles $\theta_{ij}$ without superscript refer to the PMNS matrix.

%==================================================
\section{Derivation of stable lepton mixing sum-rules\label{sumrulederivation}}
%==================================================
Let us recapitulate here the derivation of the sum-rules of our interest along the lines they were originally obtained in \cite{sumrule}.
We shall for the moment forget about the  canonical normalisation effects and drop all the hats in what follows. Later on, we shall reiterate the same procedure carefully with all the potential sources of deviations due to canonical normalisation taken into account.
 
Perhaps the simplest method to obtain (\ref{reactoranglelifting}) and (\ref{mainsumrule}) consists in looking at particular elements of the lepton mixing matrix\footnote{Since we shall be looking on the {\it magnitude} of the matrix elements, the particular phase convention employed here is immaterial, but clearly must not be altered during the computation.}:  
\be\label{srbasic}
U_{PMNS}=V_{L}^{l}V_{L}^{\nu\dagger}=V_{L}^{l}U_{TB}
\ee 
and exploiting the fact that the particular shape (\ref{VLl}) of $V_{L}^{l}$ exposes unaltered the third row of the  tri-bimaximal neutrino sector mixing in $U_{TB}$ and also the 13 entry of $U_{PMNS}$ receives a particularly simple form. Indeed, one easily obtains $|(U_{PMNS})_{3i}|=|(U_{TB})_{3i}|$ that (upon employing the standard parametrisation (\ref{Eq:StandardParametrisation})) gives in particular:
%\footnote{Note that due to the particular phase convention employed on the LHS it makes sense to match only the magnitudes of the two sides.}
\be\label{31sumrule}
31\; \mathrm{entry}:\quad \left(s_{12}s_{23}-c_{12}c_{23}s_{13}e^{i\delta}\right)e^{i\alpha_{1}/2}=s_{12}^{\nu}s_{23}^{\nu} \quad \mathrm{up\; to \; a \; global\; phase},
\ee
and also: 
\be\label{13sumrule}
13\; \mathrm{entry}:\quad s_{13}e^{-i\delta}=-s_{12}^l e^{-i\rho} s_{23}^{\nu} \quad \mathrm{up\; to \; a \; global\; phase}.
\ee 
First, notice that a nonzero $\theta_{13}$ mixing is generated from a conspiracy between the 12 charged lepton sector mixing and $\theta_{23}^{\nu}$. In a wide class of models with a built-in Georgi-Jarlskog mechanism (leading typically to $\theta_{12}^{l}\approx \theta_{C}/3$ with $\theta_{C}\approx \lambda\approx 0.2$ denoting the Cabibbo CKM mixing governed by the down-type quark sector) one gets \cite{Antusch:2005kw}
$
\theta_{13}\approx \frac{\theta_{C}}{3\sqrt{2}}
$.
Second, formula (\ref{31sumrule}) subsequently leads to
\be\label{13sumrule2}
s_{12}s_{23}-c_{12}c_{23}s_{13}\cos\delta=s_{12}^{\nu}s_{23}^{\nu}\;.
%\quad \mathrm{and}\quad s_{13}=s_{12}^l s_{23}^{\nu}
\ee
Since the 23 sector mixing is stable under the perturbation (\ref{VLl}), one can trade $s_{23}$ and $c_{23}$ in (\ref{13sumrule2}) for their TB values $\frac{1}{\sqrt{2}}$ while the RHS gives $\frac{1}{\sqrt{6}}$. Expanding the left-hand side of (\ref{13sumrule2}) for small $s_{13}\approx\theta_{13}$ one gets:  
\be\label{trading}
%s_{12}\frac{1}{\sqrt{2}}-c_{12}\frac{1}{\sqrt{2}}s_{13}\cos\delta= \frac{1}{\sqrt{6}},\quad \mathrm{i.e.}\quad
s_{12}-c_{12}\theta_{13}\cos\delta= \frac{1}{\sqrt{3}}\;.
\ee
The last step is to expand the physical $\theta_{12}$ around the tri-bimaximal value $\theta_{12}=\theta_{12}^{TB}+\Delta\theta_{12}$ which yields
$s_{12}=\frac{1}{\sqrt{3}}+\sqrt{\frac{2}{{3}}}\Delta \theta_{12}$ and 
$c_{12}=\sqrt{\frac{2}{{3}}}-\frac{1}{\sqrt{3}}\Delta \theta_{12}$, leading to:
\be\label{almostthere}
\sqrt{\frac{2}{{3}}}\Delta \theta_{12}-\theta_{13}\left(\sqrt{\frac{2}{{3}}}-\frac{1}{\sqrt{3}}\Delta \theta_{12}\right)\cos\delta= 0\;.
\ee
Forgetting about the doubly-suppressed $\theta_{13}\Delta \theta_{12}$ term on the LHS of (\ref{almostthere}), we get:
\be\label{deltatheta}
\Delta \theta_{12}\approx \theta_{13}\cos\delta \qquad \mathrm{yielding}\qquad
\theta_{12}=\theta_{12}^{TB}+\theta_{13}\cos\delta,
\ee 
providing a simple estimate for the deviation of the solar mixing angle $\theta_{12}$ from its tri-bimaximal value $\theta_{12}^{TB}= 35^{o}16'$ in terms of two other lepton sector measurables, namely the reactor mixing angle $\theta_{13}$ and the Dirac CP phase $\delta$.

%-----------------------------------------------------------------------
\paragraph{Corrections to the sum-rule $s=r\cos\delta$ :}\mbox{}\\
%-----------------------------------------------------------------------

Suppose now that the assumptions made above and which lead in particular to formula (\ref{deltatheta}) hold at the underlying flavour-model level, i.e.\ in the defining basis only. Thus, for sake of consistency with the notation used in the body of the manuscript, we shall re-equip all the relevant quantities therein with hats obtaining $\hat \theta_{12}=\theta_{12}^{TB}+\hat \theta_{13}\cos\hat\delta$ as only the leading order approximation to the physical (i.e.\ corrected) sum-rule, that should be written in terms of only unhatted quantities. 
The scope of this section is to see what happens once the effects of RG running and canonical normalisation are turned on.

Along similar lines as in section \ref{effectsonsumrules} one obtains first (utilising the perturbative procedure of section \ref{sectionperturbative} for $\eta=\eta^\mathrm{K}+\eta^\mathrm{RG}$), c.f. equations (\ref{firstturn})-(\ref{res2}):
\bea\label{Delta0}
\Delta (U_{PMNS})_{31}&=&\frac{\eta}{2}(U_{TB})_{31}\left[1-|(U_{TB})_{31}|^{2}\right]\;,
\eea
and thus (up to the Majorana phase associated to the $(U_{TB})_{31}$ entry): 
\bea
\Delta (U_{PMNS})_{31}&=&\frac{\eta}{2}\hat s_{12}^{\nu}\hat s_{23}^{\nu}\left[1-|\hat s_{12}^{\nu}\hat s_{23}^{\nu}|^{2}\right]\equiv \hat s_{12}^{\nu}\hat s_{23}^{\nu} \Delta \label{Delta}\;,
\eea
where
$
\Delta\equiv \frac{\eta}{2}\left(1-|\hat s_{12}^{\nu}\hat s_{23}^{\nu}|^{2}\right)=\frac{5}{12}\eta$.
Notice that due to the phase structure of the $\Delta (U_{PMNS})_{31}$ correction (\ref{Delta0}), and in particular the $(U_{TB})_{31}$ term therein, the overall phase of the RHS of (\ref{31sumrule}) derived from (\ref{srbasic}) and the phase of $\Delta (U_{PMNS})_{31}$ coincide.
This admits to write the analogue of relation (\ref{31sumrule}) (derived now from $U_{PMNS}=\hat U_{PMNS}+\Delta U_{PMNS}$) in a simple form:
\be\label{31sumruleprimed1}
\left(s_{12}s_{23}-c_{12}c_{23}s_{13}e^{i\delta}\right)e^{i\alpha_{1}/2}=\hat s_{12}^{\nu}\hat s_{23}^{\nu}(1+\Delta) \quad \mathrm{up\; to \; a \; global\; phase}\;.
\ee

The next step (leading to (\ref{trading}) in the `unperturbed' case) would be to trade the 23  rotations for their tri-bimaximal values $s_{23}=c_{23}=1/\sqrt{2}$, that is completely plausible if there were no K\"ahler or RG corrections around, because the neutrino part of the 23-sector rotation in $U_{PMNS}=V_{L}^{l}V_{L}^{\nu\dagger}$ formula (provided both $V_{L}^{\dagger}$ are written as $U_{23}U_{13}U_{12}$ along the lines of \cite{Antusch:2005kw}),  hits the small charged-lepton correction only (upon being grouped together with 23-rotation in $V_{L}^{l}$), with just a negligible effect on the resulting physical 23 lepton sector mixing. However, turning $\eta$ on,  $\theta_{23}$  becomes actually quite $\eta$-sensitive even in the simplest case (c.f. section \ref{simplestsetup}), and thus putting $s_{23}=c_{23}=1/\sqrt{2}$ is not good enough.

Rather than that, we shall exploit the information\footnote{Those results, though being obtained for zero $\hat \theta_{12}^{l}$, provide a good leading order estimate of the atmospheric mixing $\eta$-behaviour and since the error due to the nonzero charged-lepton 12-sector mixing (hitting such a corrected 23 mixing) is the same (at the leading order) as in the ``pure'' case (i.e.\ without K\"ahler effects), it can be neglected as far as one looks for the {\it deviations} from the original sum-rule (\ref{deltatheta}).} obtained in section \ref{simplestsetup} , see e.g.\ formula (\ref{correctionTBcase}), to write (at the leading order):
\be
s_{23}=\frac{1}{\sqrt{2}}(1+a)=\frac{1}{\sqrt{2}}\left(1+\frac{\eta}{4}\right)\quad 
\mathrm{and\;thus}
\quad
c_{23}=\frac{1}{\sqrt{2}}(1-a)=\frac{1}{\sqrt{2}}\left(1-\frac{\eta}{4}\right),
\ee
and from (\ref{31sumruleprimed1}) then (since $\Delta$ is real):
\be
s_{12}\frac{1}{\sqrt{2}}(1+a)-c_{12}\frac{1}{\sqrt{2}}(1-a)\theta_{13}\cos\delta=\frac{1}{\sqrt{6}}(1+\Delta)\;.
\ee
Expanding again the physical $\theta_{12}$ around the tri-bimaximal value $\theta_{12}=\theta_{12}^{TB}+\Delta\theta_{12}$, i.e.
$s_{12}=\frac{1}{\sqrt{3}}+\sqrt{\frac{2}{{3}}}\Delta \theta_{12}$ and 
$c_{12}=\sqrt{\frac{2}{{3}}}-\frac{1}{\sqrt{3}}\Delta \theta_{12}$ and neglecting the higher order terms in $a$, $\Delta\theta_{12}$ and $\theta_{13}$, one receives:
%\be
%(\frac{1}{\sqrt{3}}+\sqrt{\frac{2}{{3}}}\Delta \theta'_{12})\frac{1}{\sqrt{2}}(1+a)-\sqrt{\frac{2}{{3}}}\frac{1}{\sqrt{2}}\theta'_{13}\cos\delta'=\frac{1}{\sqrt{6}}+\Delta
%\ee
\be
\Delta \theta_{12}=\theta_{13}\cos\delta+\frac{1}{\sqrt{2}}(\Delta-a)=\theta_{13}\cos\delta+\frac{\eta}{6 \sqrt{2}}\;,
\ee
which is an analogue of formula (\ref{deltatheta}). The sum-rule with the K\"ahler corrections taken into account then reads:
\be\label{sumrulefinishedappendix}
\theta_{12}^{TB}=\theta_{12}-\theta_{13}\cos\delta-\frac{\eta}{6 \sqrt{2}}\;.
\ee
%where the latter form is obtained upon swapping the induced $\theta_{13}'$ for the underlying charged-sector 12 mixing angle via (\ref{13sumruleprimed}) (and neglecting the higher order $\eta \theta_{12}^{l}$-term). 
Notice that in the $\hat\theta_{12}^{l}\to 0$ limit (causing $\hat \theta_{13}\to 0$ and thus due to (\ref{13sumruleprimed}) also $\theta_{13}\to 0$) one indeed reveals the leading order effect (\ref{rsa}) in the solar mixing $s_{12}=\frac{1}{\sqrt{2}}(1+\tfrac{\eta}{6})$ obtained in section \ref{simplestsetup}, that in turn provides a non-trivial consistency check of relation (\ref{sumrulefinishedappendix}).


\begin{thebibliography}{10}

\bibitem{HPS}
%\cite{Harrison:2002er}
%bibitem{Harrison:2002er}
P.~F.~Harrison, D.~H.~Perkins and W.~G.~Scott,
%``Tri-bimaximal mixing and the neutrino oscillation data,''
Phys.\ Lett.\ B {\bf 530} (2002) 167
[arXiv:hep-ph/0202074];
%%CITATION = HEP-PH 0202074;%%
%\cite{Harrison:2002kp}
%\bibitem{Harrison:2002kp}
P.~F.~Harrison and W.~G.~Scott,
%``Symmetries and generalisations of tri-bimaximal neutrino mixing,''
Phys.\ Lett.\ B {\bf 535} (2002) 163
[arXiv:hep-ph/0203209];
%%CITATION = HEP-PH 0203209;%%
%\cite{Harrison:2003aw}
%\bibitem{Harrison:2003aw}
P.~F.~Harrison and W.~G.~Scott,
%``Permutation symmetry, tri-bimaximal neutrino mixing and the S3 group
%characters,''
Phys.\ Lett.\ B {\bf 557} (2003) 76
[arXiv:hep-ph/0302025];
%%CITATION = HEP-PH 0302025;%%
%\cite{Wolfenstein:1978uw}
%bibitem{Wolfenstein:1978uw}
an earlier related ansatz was proposed by: L.~Wolfenstein,
%``Oscillations Among Three Neutrino Types And CP Violation,''
Phys.\ Rev.\ D {\bf 18} (1978) 958.
%%CITATION = PHRVA,D18,958;%%


\bibitem{sumrule}
S.~F.~King,
%``Predicting neutrino parameters from SO(3) family symmetry and quark-lepton
%unification,''
JHEP {\bf 0508} (2005) 105
[arXiv:hep-ph/0506297];
I.~Masina,
  %``A maximal atmospheric mixing from a maximal CP violating phase,''
  Phys.\ Lett.\  B {\bf 633} (2006) 134
  [arXiv:hep-ph/0508031];
  %%CITATION = PHLTA,B633,134;%%
%%CITATION = HEP-PH 0506297;%%
S.~Antusch and S.~F.~King,
%   ``Charged lepton corrections to neutrino mixing angles and CP phases
  %revisited,''
  Phys.\ Lett.\ B {\bf 631} (2005) 42
  [arXiv:hep-ph/0508044];
  %%CITATION = HEP-PH 0508044;%%
S.~Antusch, P.~Huber, S.~F.~King and T.~Schwetz,
  %``Neutrino mixing sum rules and oscillation experiments,''
  JHEP {\bf 0704} (2007) 060
  [arXiv:hep-ph/0702286].
  %%CITATION = JHEPA,0704,060;%%


%\cite{Frampton:2004ud}
\bibitem{Frampton:2004ud}
  P.~H.~Frampton, S.~T.~Petcov and W.~Rodejohann,
  %``On deviations from bimaximal neutrino mixing,''
  Nucl.\ Phys.\  B {\bf 687} (2004) 31
  [arXiv:hep-ph/0401206];
  %%CITATION = NUPHA,B687,31;%%
%\cite{Dighe:2006sr}
%\bibitem{Dighe:2006sr}
  A.~Dighe, S.~Goswami and W.~Rodejohann,
  %``Corrections to Tri-bimaximal Neutrino Mixing: Renormalisation and Planck
  %Scale Effects,''
  Phys.\ Rev.\  D {\bf 75} (2007) 073023
  [arXiv:hep-ph/0612328];
  %%CITATION = PHRVA,D75,073023;%%
F.~Plentinger and W.~Rodejohann,
  %``Deviations from tribimaximal neutrino mixing,''
  Phys.\ Lett.\ B {\bf 625} (2005) 264
  [arXiv:hep-ph/0507143];
  %%CITATION = HEP-PH 0507143;%%
R.~N.~Mohapatra and W.~Rodejohann,
  %``Broken mu-tau symmetry and leptonic CP violation,''
  Phys.\ Rev.\ D {\bf 72} (2005) 053001
  [arXiv:hep-ph/0507312];
  %%CITATION = HEP-PH 0507312;%%
%\cite{Hochmuth:2007wq}
%\bibitem{Hochmuth:2007wq}
  K.~A.~Hochmuth, S.~T.~Petcov and W.~Rodejohann,
  %``U_{PMNS} = U_ell^dagger U_nu,''
  arXiv:0706.2975 [hep-ph].
  %%CITATION = ARXIV:0706.2975;%%



%\cite{Altarelli:2006kg}
\bibitem{Altarelli:2006kg}
  G.~Altarelli, F.~Feruglio and Y.~Lin,
  %``Tri-bimaximal neutrino mixing from orbifolding,''
  Nucl.\ Phys.\  B {\bf 775} (2007) 31
  [arXiv:hep-ph/0610165];
  %%CITATION = NUPHA,B775,31;%%
%\cite{Altarelli:2005yx}
%\bibitem{Altarelli:2005yx}
  G.~Altarelli and F.~Feruglio,
  %``Tri-bimaximal neutrino mixing, A(4) and the modular symmetry,''
  Nucl.\ Phys.\  B {\bf 741} (2006) 215
  [arXiv:hep-ph/0512103];
  %%CITATION = NUPHA,B741,215;%%
%\cite{Altarelli:2005yp}
%\bibitem{Altarelli:2005yp}
  G.~Altarelli and F.~Feruglio,
  %``Tri-bimaximal neutrino mixing from discrete symmetry in extra
  %dimensions,''
  Nucl.\ Phys.\  B {\bf 720} (2005) 64
  [arXiv:hep-ph/0504165];
  %%CITATION = NUPHA,B720,64;%%
%\cite{Feruglio:2007uu}
%\bibitem{Feruglio:2007uu}
  F.~Feruglio, C.~Hagedorn, Y.~Lin and L.~Merlo,
  %``Tri-bimaximal neutrino mixing and quark masses from a discrete flavour
  %symmetry,''
  Nucl.\ Phys.\  B {\bf 775} (2007) 120
  [arXiv:hep-ph/0702194].
  %%CITATION = NUPHA,B775,120;%%
%\bibitem{Luhn:2007sy}
  C.~Luhn, S.~Nasri and P.~Ramond,
  %``Tri-Bimaximal Neutrino Mixing and the Family Symmetry Z_7 \rtimes Z_3,''
  Phys.\ Lett.\  B {\bf 652} (2007) 27
  [arXiv:0706.2341 [hep-ph]].
  %%CITATION = PHLTA,B652,27;%%



%\cite{Ma:2007wu}
\bibitem{Ma:2007wu}
  E.~Ma,
  %``Near Tribimaximal Neutrino Mixing with Delta(27) Symmetry,''
  arXiv:0709.0507 [hep-ph];
  %%CITATION = ARXIV:0709.0507;%%
%\cite{Ma:2007ku}
%\bibitem{Ma:2007ku}
  E.~Ma,
  %``New lepton family symmetry and neutrino tribimaximal mixing,''
  arXiv:hep-ph/0701016;
  %%CITATION = HEP-PH/0701016;%%
%\cite{Ma:2006vq}
%\bibitem{Ma:2006vq}
  E.~Ma,
  %``Supersymmetric A(4) x Z(3) and A(4) realisations of neutrino  tribimaximal
  %mixing without and with corrections,''
  Mod.\ Phys.\ Lett.\  A {\bf 22} (2007) 101
  [arXiv:hep-ph/0610342];
  %%CITATION = MPLAE,A22,101;%%
%\cite{Ma:2006wm}
%\bibitem{Ma:2006wm}
  E.~Ma,
  %``Suitability of A(4) as a family symmetry in grand unification,''
  Mod.\ Phys.\ Lett.\  A {\bf 21} (2006) 2931
  [arXiv:hep-ph/0607190];
  %%CITATION = MPLAE,A21,2931;%%
%\cite{Ma:2006ip}
%\bibitem{Ma:2006ip}
  E.~Ma,
  %``Neutrino mass matrix from Delta(27) symmetry,''
  Mod.\ Phys.\ Lett.\  A {\bf 21} (2006) 1917
  [arXiv:hep-ph/0607056];
  %%CITATION = MPLAE,A21,1917;%%
%\cite{Ma:2006sk}
%\bibitem{Ma:2006sk}
  E.~Ma, H.~Sawanaka and M.~Tanimoto,
  %``Quark masses and mixing with A(4) family symmetry,''
  Phys.\ Lett.\  B {\bf 641} (2006) 301
  [arXiv:hep-ph/0606103];
  %%CITATION = PHLTA,B641,301;%%
%\cite{Ma:2006dn}
%\bibitem{Ma:2006dn}
  E.~Ma,
  %``Tribimaximal neutrino mixing from a supersymmetric model with A4 family
  %symmetry,''
  Phys.\ Rev.\  D {\bf 73} (2006) 057304;
  %%CITATION = PHRVA,D73,057304;%%
%\cite{Adhikary:2006wi}
%\bibitem{Adhikary:2006wi}
  B.~Adhikary, B.~Brahmachari, A.~Ghosal, E.~Ma and M.~K.~Parida,
  %``A(4) symmetry and prediction of U(e3) in a modified Altarelli-Feruglio
  %model,''
  Phys.\ Lett.\  B {\bf 638} (2006) 345
  [arXiv:hep-ph/0603059];
  %%CITATION = PHLTA,B638,345;%%
%\cite{Ma:2005mw}
%\bibitem{Ma:2005mw}
  E.~Ma,
  %``Tetrahedral family symmetry and the neutrino mixing matrix,''
  Mod.\ Phys.\ Lett.\  A {\bf 20} (2005) 2601
  [arXiv:hep-ph/0508099];
  %%CITATION = MPLAE,A20,2601;%%
%\cite{Ma:2005sha}
%\bibitem{Ma:2005sha}
  E.~Ma,
  %``Aspects of the tetrahedral neutrino mass matrix,''
  Phys.\ Rev.\  D {\bf 72} (2005) 037301
  [arXiv:hep-ph/0505209];
  %%CITATION = PHRVA,D72,037301;%%
%\cite{Chen:2005jm}
%\bibitem{Chen:2005jm}
  S.~L.~Chen, M.~Frigerio and E.~Ma,
  %``Hybrid seesaw neutrino masses with A(4) family symmetry,''
  Nucl.\ Phys.\  B {\bf 724} (2005) 423
  [arXiv:hep-ph/0504181];
  %%CITATION = NUPHA,B724,423;%%
%\cite{Ma:2004zv}
%\bibitem{Ma:2004zv}
  E.~Ma,
  %``A(4) origin of the neutrino mass matrix,''
  Phys.\ Rev.\  D {\bf 70} (2004) 031901
  [arXiv:hep-ph/0404199].
  %%CITATION = PHRVA,D70,031901;%%




%\cite{deMedeirosVarzielas:2005ax}
\bibitem{deMedeirosVarzielas:2005ax}
  I.~de Medeiros Varzielas and G.~G.~Ross,
  %``SU(3) family symmetry and neutrino bi-tri-maximal mixing,''
  Nucl.\ Phys.\  B {\bf 733} (2006) 31
  [arXiv:hep-ph/0507176];
  %%CITATION = NUPHA,B733,31;%%
%\cite{deMedeirosVarzielas:2005qg}
%\bibitem{deMedeirosVarzielas:2005qg}
  I.~de Medeiros Varzielas, S.~F.~King and G.~G.~Ross,
  %``Tri-bimaximal neutrino mixing from discrete subgroups of SU(3) and  SO(3)
  %family symmetry,''
  Phys.\ Lett.\  B {\bf 644} (2007) 153
  [arXiv:hep-ph/0512313];
  %%CITATION = PHLTA,B644,153;%%
%\cite{deMedeirosVarzielas:2006fc}
%\bibitem{deMedeirosVarzielas:2006fc}
  I.~de Medeiros Varzielas, S.~F.~King and G.~G.~Ross,
  %``Neutrino tri-bi-maximal mixing from a non-Abelian discrete family
  %symmetry,''
  Phys.\ Lett.\  B {\bf 648} (2007) 201
  [arXiv:hep-ph/0607045];
  %%CITATION = PHLTA,B648,201;%%

%\cite{King:2006np}
\bibitem{King:2006np}
  S.~F.~King and M.~Malinsky,
  %``A(4) family symmetry and quark-lepton unification,''
  Phys.\ Lett.\  B {\bf 645} (2007) 351
  [arXiv:hep-ph/0610250];
  %%CITATION = PHLTA,B645,351;%%
%\cite{King:2006me}
%\bibitem{King:2006me}
  S.~F.~King and M.~Malinsky,
  %``Towards a complete theory of fermion masses and mixings with SO(3)  family
  %symmetry and 5d SO(10) unification,''
  JHEP {\bf 0611} (2006) 071
  [arXiv:hep-ph/0608021];
  %%CITATION = JHEPA,0611,071;%%
%\cite{Luhn:2007sy}



%\cite{Harrison:2003aw}
\bibitem{Harrison:2003aw}
  P.~F.~Harrison and W.~G.~Scott,
  %``Permutation symmetry, tri-bimaximal neutrino mixing and the S3 group
  %characters,''
  Phys.\ Lett.\  B {\bf 557} (2003) 76
  [arXiv:hep-ph/0302025];
  %%CITATION = PHLTA,B557,76;%%
%\cite{Harrison:2002kp}
%\bibitem{Harrison:2002kp}
  P.~F.~Harrison and W.~G.~Scott,
  %``Symmetries and generalisations of tri-bimaximal neutrino mixing,''
  Phys.\ Lett.\  B {\bf 535} (2002) 163
  [arXiv:hep-ph/0203209];
  %%CITATION = PHLTA,B535,163;%%
%\cite{Mohapatra:2006pu}
%\bibitem{Mohapatra:2006pu}
  R.~N.~Mohapatra, S.~Nasri and H.~B.~Yu,
  %``S(3) symmetry and tri-bimaximal mixing,''
  Phys.\ Lett.\  B {\bf 639} (2006) 318
  [arXiv:hep-ph/0605020];
  %%CITATION = PHLTA,B639,318;%%
%\cite{Mohapatra:2006se}
%\bibitem{Mohapatra:2006se}
  R.~N.~Mohapatra and H.~B.~Yu,
  %``Connecting leptogenesis to CP violation in neutrino mixings in a
  %tri-bimaximal mixing model,''
  Phys.\ Lett.\  B {\bf 644} (2007) 346
  [arXiv:hep-ph/0610023];
  %%CITATION = PHLTA,B644,346;%%
%\cite{Chen:2007af}
%\bibitem{Chen:2007af}
  M.~C.~Chen and K.~T.~Mahanthappa,
  %``CKM and Tri-bimaximal MNS Matrices in a SU(5) x (d)T Model,''
  Phys.\ Lett.\  B {\bf 652} (2007) 34
  [arXiv:0705.0714 [hep-ph]];
  %%CITATION = PHLTA,B652,34;%%
%\cite{Low:2003dz}
%\bibitem{Low:2003dz}
  C.~I.~Low and R.~R.~Volkas,
  %``Tri-bimaximal mixing, discrete family symmetries, and a conjecture
  %connecting the quark and lepton mixing matrices,''
  Phys.\ Rev.\  D {\bf 68} (2003) 033007
  [arXiv:hep-ph/0305243];
  %%CITATION = PHRVA,D68,033007;%%
%\cite{He:2006et}
%\bibitem{He:2006et}
  X.~G.~He,
  %``A(4) group and tri-bimaximal neutrino mixing: A renormalizable model,''
  Nucl.\ Phys.\ Proc.\ Suppl.\  {\bf 168} (2007) 350
  [arXiv:hep-ph/0612080];
  %%CITATION = NUPHZ,168,350;%%
%\cite{Aranda:2007dp}
%\bibitem{Aranda:2007dp}
  A.~Aranda,
  %``Neutrino mixing from the double tetrahedral group $T^{\prime}$,''
  arXiv:0707.3661 [hep-ph].
  %%CITATION = ARXIV:0707.3661;%%


%\cite{Chan:2007ng}
\bibitem{Chan:2007ng}
  A.~H.~Chan, H.~Fritzsch and Z.~z.~Xing,
  %``Deviations from Tri-bimaximal Neutrino Mixing in Type-II Seesaw and
  %Leptogenesis,''
  arXiv:0704.3153 [hep-ph];
  Z.~z.~Xing,
  %``Nontrivial correlation between the CKM and MNS matrices,''
  Phys.\ Lett.\ B {\bf 618} (2005) 141
  [arXiv:hep-ph/0503200];
  %%CITATION = HEP-PH 0503200;%%
%\cite{Xing:2006xa}
%\bibitem{Xing:2006xa}
  Z.~z.~Xing, H.~Zhang and S.~Zhou,
  %``Nearly tri-bimaximal neutrino mixing and CP violation from mu - tau
  %symmetry breaking,''
  Phys.\ Lett.\  B {\bf 641} (2006) 189
  [arXiv:hep-ph/0607091];
  %%CITATION = PHLTA,B641,189;%%
%\cite{Kang:2005bg}
%\bibitem{Kang:2005bg}
  S.~K.~Kang, Z.~z.~Xing and S.~Zhou,
  %``Possible deviation from the tri-bimaximal neutrino mixing in a seesaw
  %model,''
  Phys.\ Rev.\  D {\bf 73} (2006) 013001
  [arXiv:hep-ph/0511157];
  %%CITATION = PHRVA,D73,013001;%%
%\cite{Luo:2005fc}
%\bibitem{Luo:2005fc}
  S.~Luo and Z.~z.~Xing,
  %``Generalized tri-bimaximal neutrino mixing and its sensitivity to  radiative
  %corrections,''
  Phys.\ Lett.\  B {\bf 632} (2006) 341
  [arXiv:hep-ph/0509065];
  %%CITATION = PHLTA,B632,341;%%
%\cite{Hirsch:2006je}
%\bibitem{Hirsch:2006je}
  M.~Hirsch, E.~Ma, J.~C.~Romao, J.~W.~F.~Valle and A.~Villanova del Moral,
  %``Minimal supergravity radiative effects on the tri-bimaximal neutrino
  %mixing pattern,''
  Phys.\ Rev.\  D {\bf 75} (2007) 053006
  [arXiv:hep-ph/0606082];
  %%CITATION = PHRVA,D75,053006;%%
%\cite{Singh:2006dr}
%\bibitem{Singh:2006dr}
  N.~N.~Singh, M.~Rajkhowa and A.~Borah,
  %``Deviation from tri-bimaximal mixings in two types of inverted  hierarchical
  %neutrino mass models,''
  arXiv:hep-ph/0603189;
  %%CITATION = HEP-PH/0603189;%%
%\cite{He:2006qd}
%\bibitem{He:2006qd}
  X.~G.~He and A.~Zee,
  %``Minimal modification to the tri-bimaximal neutrino mixing,''
  Phys.\ Lett.\  B {\bf 645} (2007) 427
  [arXiv:hep-ph/0607163];
  %%CITATION = PHLTA,B645,427;%
%\cite{Haba:2006dz}
%\bibitem{Haba:2006dz}
  N.~Haba, A.~Watanabe and K.~Yoshioka,
  %``Twisted flavors and tri/bi-maximal neutrino mixing,''
  Phys.\ Rev.\ Lett.\  {\bf 97} (2006) 041601
  [arXiv:hep-ph/0603116].
  %%CITATION = PRLTA,97,041601;%%

%\cite{Antusch:2007jd}
\bibitem{Antusch:2007jd}
  S.~Antusch, L.~E.~Ibanez and T.~Macri,
  %``Neutrino Masses and Mixings from String Theory Instantons,''
  JHEP {\bf 0709} (2007) 087.
%  [arXiv:0706.2132 [hep-ph]].
  %%CITATION = JHEPA,0709,087;%%
  %\cite{King:2006hn}


\bibitem{Antusch:2007re}
S.~Antusch, S.~F. King, and M.~Malinsky, 
%{\it Solving the susy flavour and cp
 % problems with su(3) family symmetry},
  \href{http://xxx.lanl.gov/abs/arXiv:0708.1282 [hep-ph]}{{\tt arXiv:0708.1282
  [hep-ph]}}.


%\cite{Leurer:1993gy}
\bibitem{Leurer:1993gy}
  M.~Leurer, Y.~Nir and N.~Seiberg,
  %``Mass matrix models: The Sequel,''
  Nucl.\ Phys.\  B {\bf 420} (1994) 468
  [arXiv:hep-ph/9310320];
  %%CITATION = NUPHA,B420,468;%%
%\bibitem{dudas} E.~Dudas, S.~Pokorski and C.~A.~Savoy, 
%``Yukawa matrices from a spontaneously broken Abelian symmetry,''
Phys.\ Lett.\ B \textbf{356} (1995) 45 [arXiv:hep-ph/9504292]; 
%%CITATION = HEP-PH 9504292;%%
\newline
E.~Dudas, S.~Pokorski and C.~A.~Savoy, 
%``Soft scalar masses in supergravity with horizontal U(1)_X gauge symmetry,''
Phys.\ Lett.\ B \textbf{369} (1996) 255 [arXiv:hep-ph/9509410]; 
%%CITATION = HEP-PH 9509410;%%
%\bibitem{dreiner} H.~K.~Dreiner and M.~Thormeier, 
%``Supersymmetric Froggatt-Nielsen models with baryon- and lepton-number
%violation,''
Phys.\ Rev.\ D \textbf{69} (2004) 053002 [arXiv:hep-ph/0305270];
%%CITATION = HEP-PH 0305270;%%
%\bibitem{timjones} I.~Jack, D.~R.~T.~Jones and R.~Wild, 
%``Yukawa textures and the mu-term,''
Phys.\ Lett.\ B \textbf{580} (2004) 72 [arXiv:hep-ph/0309165].
%%CITATION = HEP-PH 0309165;%%


%\cite{King:2003xq}
\bibitem{King:2003xq}
  S.~F.~King and I.~N.~R.~Peddie,
  %``Canonical normalisation and Yukawa matrices,''
  Phys.\ Lett.\  B {\bf 586} (2004) 83
  [arXiv:hep-ph/0312237].
  %%CITATION = PHLTA,B586,83;%%

%\cite{King:2004tx}
\bibitem{King:2004tx}
  S.~F.~King, I.~N.~R.~Peddie, G.~G.~Ross, L.~Velasco-Sevilla and O.~Vives,
  %``Kaehler corrections and softly broken family symmetries,''
  JHEP {\bf 0507} (2005) 049
  [arXiv:hep-ph/0407012].
  %%CITATION = JHEPA,0507,049;%%

%\cite{Antusch:2007ib}
\bibitem{Antusch:2007ib}
  S.~Antusch, S.~F.~King and M.~Malinsky,
  %``Third Family Corrections to Tri-bimaximal Lepton Mixing and a New Sum
  %Rule,''
  arXiv:0711.4727 [hep-ph].
  %%CITATION = ARXIV:0711.4727;%%

%\cite{Group:2007kx}
\bibitem{Group:2007kx}
 The ISS Physics Working Group,
  ``Physics at a future Neutrino Factory and super-beam facility,''
  arXiv:0710.4947 [hep-ph].
  %%CITATION = ARXIV:0710.4947;%%

\bibitem{seesaw}
P.~Minkowski,
  %``Mu $\to$ E Gamma At A Rate Of One Out Of 1-Billion Muon Decays?,''
  Phys.\ Lett.\ B {\bf 67} (1977) 421;
  %%CITATION = PHLTA,B67,421;%%
M. Gell-Mann, P. Ramond and R. Slansky in Sanibel Talk,
CALT-68-709, Feb 1979, and in {\it Supergravity} (North Holland,
Amsterdam 1979);
T. Yanagida in {\it Proc. of the Workshop on Unified Theory and
Baryon Number of the Universe}, KEK, Japan, 1979;
S.L.Glashow, Cargese Lectures (1979);
%\cite{Mohapatra:1979ia}
%\bibitem{Mohapatra:1979ia}
R.~N.~Mohapatra and G.~Senjanovic,
%``Neutrino Mass And Spontaneous Parity Nonconservation,''
Phys.\ Rev.\ Lett.\  {\bf 44} (1980) 912;
%%CITATION = PRLTA,44,912;%%
J.~Schechter and J.~W.~Valle,
%``Neutrino Decay And Spontaneous Violation Of Lepton Number,''
Phys.\ Rev.\ D {\bf 25} (1982) 774.
%%CITATION = PHRVA,D25,774;%%

\bibitem{Yao:2006px}
{\bf Particle Data Group} Collaboration, W.~M. Yao {\em et~al.}, {\it Review of
  particle physics},  {\em J. Phys.} {\bf G33} (2006) 1--1232.

%\cite{deMedeirosVarzielas:2006fc}
\bibitem{deMedeirosVarzielas:2006fc}
  I.~de Medeiros Varzielas, S.~F.~King and G.~G.~Ross,
  %``Neutrino tri-bi-maximal mixing from a non-Abelian discrete family
  %symmetry,''
  Phys.\ Lett.\  B {\bf 648} (2007) 201
  [arXiv:hep-ph/0607045].
  %%CITATION = PHLTA,B648,201;%%

%\cite{Malinsky:2007pf}
\bibitem{Malinsky:2007pf}
  M.~Malinsky,
  %``Tackling the SUSY flavour & CP problem - SUGRA versus SU(3),''
  arXiv:0710.2430 [hep-ph].
  %%CITATION = ARXIV:0710.2430;%%


\bibitem{King:2006hn}
The precise meaning of the statement that the TB mixing comes entirely
from the neutrino sector is discussed for example in:
  S.~F.~King,
  %``Invariant see-saw models and sequential dominance,''
  Nucl.\ Phys.\  B {\bf 786} (2007) 52.
%  [arXiv:hep-ph/0610239].
  %%CITATION = NUPHA,B786,52;%%

%\cite{King:2007pr}
\bibitem{King:2007pr}
  S.~F.~King,
  %``Parametrizing the lepton mixing matrix in terms of deviations from
  %tri-bimaximal mixing,''
  arXiv:0710.0530 [hep-ph].
  %%CITATION = ARXIV:0710.0530;%%

%\cite{King:1998jw}
\bibitem{King:1998jw}
S.~F.~King,
%``Atmospheric and solar neutrinos with a heavy singlet,''
Phys.\ Lett.\ B {\bf 439} (1998) 350
[arXiv:hep-ph/9806440];
%%CITATION = HEP-PH 9806440;%%
S.~F.~King,
%``Atmospheric and solar neutrinos from single right-handed neutrino  dominance
%and U(1) family symmetry,''
Nucl.\ Phys.\ B {\bf 562} (1999) 57
[arXiv:hep-ph/9904210];
%%CITATION = HEP-PH 9904210;%%
S.~F.~King,
%``Large mixing angle MSW and atmospheric neutrinos from single  right-handed
%neutrino dominance and U(1) family symmetry,''
Nucl.\ Phys.\ B {\bf 576} (2000) 85
[arXiv:hep-ph/9912492];
%\cite{King:2002nf}
%\bibitem{King:2002nf}
S.~F.~King,
%``Constructing the large mixing angle MNS matrix in see-saw models with
%right-handed neutrino dominance,''
JHEP {\bf 0209} (2002) 011
[arXiv:hep-ph/0204360].
%%CITATION = HEP-PH 0204360;%%
For a review, see:  
  S.~Antusch and S.~F.~King,
  %``Sequential dominance,''
  New J.\ Phys.\  {\bf 6} (2004) 110
  [arXiv:hep-ph/0405272].
  %%CITATION = HEP-PH 0405272;%%

\bibitem{SU(3)messenger}
S.~F.~King and G.~G.~Ross,
  %``Fermion masses and mixing angles from SU(3) family symmetry,''
  Phys.\ Lett.\  B {\bf 520} (2001) 243
  [arXiv:hep-ph/0108112];
  %%CITATION = PHLTA,B520,243;%%
S.~F.~King and G.~G.~Ross,
  %``Fermion masses and mixing angles from SU(3) family symmetry and
  %unification,''
  Phys.\ Lett.\  B {\bf 574} (2003) 239
  [arXiv:hep-ph/0307190].
  %%CITATION = PHLTA,B574,239;%%

\bibitem{RGE}
P.~H. Chankowski and Z.~Pluciennik, 
%\emph{Renormalisation group equations for
%  seesaw neutrino masses}, 
Phys. Lett. \textbf{B316} (1993), 312--317,
  \texttt{hep-ph/9306333};
%
%\bibitem{Babu:1993qv}
K.~S. Babu, C.~N. Leung, and J.~Pantaleone, 
%\emph{Renormalization of the
%  neutrino mass operator}, Phys. Lett. \textbf{B319} (1993), 191--198,
  \texttt{hep-ph/9309223};
S.~F.~King and N.~N.~Singh,
  %``Renormalisation group analysis of single right-handed neutrino
  %dominance,''
  Nucl.\ Phys.\  B {\bf 591} (2000) 3, \texttt{hep-ph/0006229};
%\bibitem{Antusch:2001ck}
S.~Antusch, M.~Drees, J.~Kersten, M.~Lindner, and M.~Ratz, 
%\emph{Neutrino mass
%  operator renormalization revisited}, 
Phys. Lett. \textbf{B519} (2001),
  238--242,  \texttt{hep-ph/0108005};
%
%\bibitem{Antusch:2001vn}
S.~Antusch, M.~Drees, J.~Kersten, M.~Lindner, and M.~Ratz, 
%\emph{Neutrino mass
%  operator renormalization in {T}wo {H}iggs {D}oublet {M}odels and the {MSSM}},
Phys.\ Lett.\  B {\bf 525} (2002) 130
  [arXiv:hep-ph/0110366];
  %%CITATION = PHLTA,B525,130;%%
%
S.~Antusch, J.~Kersten, M.~Lindner, and M.~Ratz, 
%\emph{Neutrino mass matrix
%  running for non-degenerate see-saw scales}, 
Phys. Lett. \textbf{B538} (2002),  87--95,  \texttt{hep-ph/0203233};
%
%\bibitem{Antusch:2002ek}
S.~Antusch and M.~Ratz, 
%\emph{Supergraph techniques and two-loop 
%beta-functions
%  for renormalizable and non-renormalizable operators}, 
JHEP \textbf{07}  (2002), 059,  \texttt{hep-ph/0203027}.


\bibitem{RGEanalytical}
%\bibitem{Antusch:2005gp}
  S.~Antusch, J.~Kersten, M.~Lindner, M.~Ratz and M.~A.~Schmidt,
  %``Running neutrino mass parameters in see-saw scenarios,''
  JHEP {\bf 0503} (2005) 024
  [arXiv:hep-ph/0501272].
%\bibitem{Antusch:2003kp}
  S.~Antusch, J.~Kersten, M.~Lindner and M.~Ratz,
  %``Running neutrino masses, mixings and CP phases: Analytical results and
  %phenomenological consequences,''
  Nucl.\ Phys.\  B {\bf 674} (2003) 401
  [arXiv:hep-ph/0305273].
  
  
\bibitem{Dighe:2006sr}
  For a recent analysis, see: A.~Dighe, S.~Goswami and W.~Rodejohann,
  %``Corrections to Tri-bimaximal Neutrino Mixing: Renormalization and Planck
  %Scale Effects,''
  Phys.\ Rev.\  D {\bf 75} (2007) 073023
  [arXiv:hep-ph/0612328];
  %%CITATION = PHRVA,D75,073023;%%
A.~Dighe, S.~Goswami and P.~Roy,
  %``Radiatively broken symmetries of nonhierarchical neutrinos,''
  Phys.\ Rev.\  D {\bf 76} (2007) 096005
  [arXiv:0704.3735 [hep-ph]].
  %%CITATION = PHRVA,D76,096005;%%


%\cite{Antusch:2005gp}
\bibitem{Antusch:2005gp}
  S.~Antusch, J.~Kersten, M.~Lindner, M.~Ratz and M.~A.~Schmidt,
  %``Running neutrino mass parameters in see-saw scenarios,''
  JHEP {\bf 0503} (2005) 024
  [arXiv:hep-ph/0501272].
  %%CITATION = JHEPA,0503,024;%%

\bibitem{Antusch:2004xd}
  S.~Antusch and S.~F.~King,
  %``From hierarchical to partially degenerate neutrinos via type II upgrade  of
  %type I see-saw models,''
  Nucl.\ Phys.\  B {\bf 705} (2005) 239
  [arXiv:hep-ph/0402121].
  %%CITATION = NUPHA,B705,239;%%

\bibitem{King:2005bj}
S.~F.~King,
%``Predicting neutrino parameters from SO(3) family symmetry and quark-lepton
%unification,''
JHEP {\bf 0508} (2005) 105
[arXiv:hep-ph/0506297].

  
\bibitem{PS}
J.~C. Pati and A.~Salam, %\emph{Unified lepton - hadron symmetry and a gauge
  %theory of the basic interactions}, 
  Phys. Rev. \textbf{D8} (1973), 1240.

\bibitem{SO10}
H.~Georgi, \emph{Particles and fields}, (edited by Carlson, C. E.), A.I.P.,
  1975, p. 575;
H.~Fritzsch and P.~Minkowski, 
%\emph{Unified interactions of leptons and hadrons}, 
Ann. Phys. \textbf{93} (1975), 193--266.

\bibitem{Georgi:1979df}
  H.~Georgi and C.~Jarlskog,
  %``A New Lepton - Quark Mass Relation In A Unified Theory,''
  Phys.\ Lett.\ B {\bf 86} (1979) 297.
  %%CITATION = PHLTA,B86,297;%%
    
%\cite{Antusch:2004yx}
\bibitem{Antusch:2004yx}
  %%CITATION = PHRVA,D70,097302;%%
  S.~Antusch, P.~Huber, J.~Kersten, T.~Schwetz and W.~Winter,
  %``Is there maximal mixing in the lepton sector?,''
  Phys.\ Rev.\  D {\bf 70} (2004) 097302.
%  [arXiv:hep-ph/0404268].
 
\bibitem{Antusch:2005kw}
  S.~Antusch and S.~F.~King,
  %``Charged lepton corrections to neutrino mixing angles and CP phases
  %revisited,''
  Phys.\ Lett.\  B {\bf 631} (2005) 42
  [arXiv:hep-ph/0508044].
  %%CITATION = PHLTA,B631,42;%%

\end{thebibliography}
\end{document}